\documentclass[10pt,floats,floatfix,amssymb,aps,prd,twocolumn,superscriptaddress,nofootinbib,preprintnumbers]{revtex4-2}
		
\usepackage{amssymb,amsmath,verbatim,mathtools,needspace,enumitem,etoolbox,graphicx,physics,microtype,afterpage,bm}

\usepackage[utf8]{inputenc}
\usepackage{amsmath,amssymb}
\usepackage{amsfonts}
\usepackage{bm}
\usepackage{courier}
\usepackage{graphics,orcidlink}
\usepackage{amsmath}
\usepackage{amssymb}
\usepackage[normalem]{ulem} 
\usepackage[mathscr]{euscript}
\usepackage{acronym}
\usepackage[caption=false]{subfig}
\usepackage{lipsum}
\usepackage{mathtools}
\usepackage{multirow}
\usepackage{graphicx}
\usepackage{mathtools}
\usepackage{multirow}
\usepackage{graphicx}
\usepackage{booktabs} 
\usepackage{comment}

\usepackage{ragged2e}
\DeclareCaptionJustification{justified}{\justifying}
\makeatletter
\newcommand{\subsetsim}{\mathrel{\mathpalette\subset@sim\relax}}
\newcommand{\subset@sim}[2]{%
  \vtop{\offinterlineskip\m@th
    \ialign{\hfil##\cr
      $#1\subset$\cr\noalign{\kern0.5pt}\scalebox{0.9}{$#1\sim$}\cr
    }%
  }%
}
\makeatother

\usepackage{amssymb,amsmath,verbatim,mathtools,needspace,enumitem,etoolbox,graphicx,physics,microtype,afterpage,bm}

\usepackage{multirow}
\usepackage{dcolumn}
\usepackage{pifont}
\usepackage{lmodern}

\usepackage{multirow}

\allowdisplaybreaks
\usepackage{tikz}
\usepackage{color}
\usepackage{framed}
\usepackage{hyperref}
\usepackage[all]{hypcap}
\definecolor{rossos}{cmyk}{0,1,1,0.55}
\definecolor{bluscuro}{rgb}{0.15, 0.2, .85}
\definecolor{bluchiaro}{cmyk}{1,.3,0.,0.1}
\definecolor{ForestGreen}{rgb}{0.13, 0.55, 0.13}
\definecolor{azure}{rgb}{0.0, 0.5, 1.0}
 
\usepackage{slashed}
\usepackage{hyperref}
\usepackage{tikz}
\usepackage[dvipsnames,svgnames,x11names]{xcolor}
\usepackage[capitalise]{cleveref}
\usetikzlibrary{arrows.meta,decorations.pathmorphing,decorations.markings}
\usetikzlibrary{calc}

\newcommand{\be}{\begin{equation}}
\newcommand{\ee}{\end{equation}}

\def\lsim{\mathrel{\rlap{\lower4pt\hbox{\hskip0.5pt$\sim$}}
    \raise1pt\hbox{$<$}}}         
\def\gsim{\mathrel{\rlap{\lower4pt\hbox{\hskip0.5pt$\sim$}}
    \raise1pt\hbox{$>$}}}         

\def\R{\mathcal{R}}
\def\S{\mathcal{S}}

\def\ttau{\tilde \tau}
\def\tvarphi{\tilde \varphi}
\def\tvarphi{\tilde \varphi}
\def\tu{\tilde u}

\usepackage{tcolorbox} 

\tcbset{
  blackbox/.style={
    colframe=black,     
    colback=white,    
    sharp corners,    
    boxrule=0.2mm       
  }
}

\makeatletter
\def\l@subsubsection#1#2{}
\makeatother

\newcommand{\nottingham}{Nottingham Centre of Gravity \& School of Mathematical Sciences, University of Nottingham, University Park, Nottingham, NG7 2RD, UK}
\newcommand{\GSSI}{Gran Sasso Science Institute (GSSI), I-67100 L'Aquila, Italy and INFN, Laboratori Nazionali del Gran Sasso, I-67100 Assergi, Italy}
\newcommand{\dublin}{School of Mathematics \& Statistics, University College Dublin, Belfield, Dublin 4, Ireland, D04 V1W8}

\begin{document}
\title{
Dynamics of a spinning body with scalar multipoles
}

\begin{abstract}

We construct an effective worldline action for a spinning compact object in theories of gravity with massless, non-minimally coupled scalar fields. To this end, we generalize the standard skeletonized approach used in General Relativity to model the stress-energy tensor of a small body in a background spacetime, dressing its multipolar structure with the scalar field. Within a pole-dipole approximation, we derive a modified set of Mathisson-Papapetrou-Dixon equations of motion for the spinning body, together with the field equations for the scalar field and the metric. We consider both parity-even and parity-odd scalars.
%
We tailor this formalism to model the evolution of asymmetric binaries, such as Extreme Mass Ratio Inspirals (EMRIs), by systematically expanding the relevant equations in powers of the binary mass ratio and deriving the dynamics up to first post-adiabatic (1PA) order, along with the self-force equations needed to determine the particle’s acceleration. For a parity-even scalar, the 1PA dynamics depends only on the mass, spin, and scalar charge of the secondary, plus a single self-coupling constant. For a pseudoscalar, we find that the binary dynamics coincides with General Relativity through 1PA.
\end{abstract}

\author{Jacopo Lestingi}
 \email{jacopo.lestingi@nottingham.ac.uk}
\affiliation{\nottingham}

\author{Thomas P. Sotiriou}
\affiliation{\nottingham}
\affiliation{School of Physics and Astronomy, University of Nottingham, University Park, Nottingham, NG7 2RD, UK}

\author{Andrew Spiers}
\affiliation{\dublin}

\author{Andrea Maselli}
\affiliation{\GSSI}

\date{\today}
\maketitle

  \hypersetup{linkcolor=black}

\begingroup
  \hypersetup{hidelinks}
  \tableofcontents
\endgroup

\section{Introduction}\label{sec:intro}

The notion of the point particle is ubiquitous 
in science, standing as a building block of 
classical physics, where it offers a faithful 
representation of real, extended sources across 
a breadth of phenomena related to mechanics 
and electromagnetism. In fact, even 
non-linear phenomena, such as the interaction 
of a charge distribution with its own 
electromagnetic field --- the Abraham--Lorentz 
self-force --- can be rigorously treated by 
taking the limit in which the size of the 
distribution tends to zero~\cite{Gralla:2009md}.

The concept of point particles, however, becomes 
more precarious in General Relativity~(GR), where 
it inevitably fails due to the intrinsic nonlinearity 
of the Einstein equations~\cite{Gralla:2008fg,Poisson:2011nh}. Pioneering 
work by Geroch and Traschen showed that a point 
particle is too singular for its curvature to be 
well defined within the theory of distributions~\cite{Geroch:1986jjl}.

While this represents a mathematical no-go 
statement, point particles are still widely 
used in GR, as it turns out that their 
generalization to extended, nonlinear matter 
distributions follows a well-defined path, 
resting on two pillars: a systematic multipole 
decomposition of the source, and a 
regularization scheme to handle the singularities 
inherent to the point-particle limit~\cite{Barack:2018yvs,Musaeus:2026kgp}.

The standard lore traces back to early works by Mathisson, Dixon and 
Tulczyjew~\cite{Mathisson2010, Dixon:1974xoz, Tulczyew}, which build upon the concept of the 
\textit{skeletonized body}, in which a small object embedded in a 
background geometry produces a small perturbation of the latter. For a 
sufficiently well-behaved matter distribution, the source of such 
perturbation --- the object's stress-energy tensor --- can be reduced 
to an infinite set of multipole moments,
\begin{equation}
T^{\mu\nu} \sim \int d\tau\, \Bigl\{
  p^{(\mu}u^{\nu)} \delta^{4}
  - \nabla_\rho\bigl[S^{\rho(\mu} u^{\nu)}\, \delta^{4}\bigr] 
  + \cdots
\Bigr\}\,,
\label{eq:Tmunu_multipoles}
\end{equation}
where $\delta^{4} = \frac{1}{\sqrt{-g}}\delta^{(4)}(x^\alpha - z^\alpha_p)$ 
is the covariant Dirac distribution evaluated on the worldline $\gamma$ 
with coordinates $z^\alpha_p$ and 4-velocity $u^\mu$. The coefficients 
$m$ and $S^{\mu\nu} = S^{[\mu\nu]}$ correspond to the body's mass and spin, 
while higher derivatives of $\delta^4$ couple to the quadrupole 
and higher moments. Eq.~\eqref{eq:Tmunu_multipoles} provides a clean 
way to treat structured compact objects, with its leading term 
identifying the structureless point-particle limit.

The source's multipole moments provide the key ingredients for the 
theory of \textit{matched} asymptotic expansions, which has become 
the standard framework for deriving the equations of motion of small 
objects in physical problems characterized by a hierarchy of 
scales~\cite{Eckhaus1979,Kevorkian1996}. The core of this approach rests on the existence of an inner 
and an outer zone, in which gravity is dominated by the small object, 
of size $\ell$, and by the background spacetime $\bar{g}_{\mu\nu}$, of 
lengthscale ${\cal L}\gg \ell$, respectively. The dynamics of the 
small body is determined by requiring that the solutions to the field 
equations in both zones match in a buffer region, where the influence 
of the body and of the background are comparable. It turns out that 
the field in the inner zone, evaluated at distances $r \gg \ell$ from 
the body, admits a natural multipole expansion, through which the 
moments emerge as the complete encoding of the body's relevant 
physical properties, and which source the perturbations of the outer 
problem. The core result of the procedure is an expansion of the full 
spacetime metric, $g_{\mu\nu}=\bar{g}_{\mu\nu}+\epsilon h^{(1)}_{\mu\nu}+\epsilon^2 
h^{(2)}_{\mu\nu}+\cdots$, where the perturbations $h^{(i)}_{\mu\nu}$ are 
sourced by the small body, and $\epsilon=\ell/{\cal L}\ll 1$.

The method of matched asymptotic expansions~\cite{Kevorkian1996,fenichel1979geometric,Eckhaus1979} has been consecrated as 
a keystone of gravitational self-force theory, which constitutes a 
rigorous framework for describing the dynamics of small bodies 
orbiting in a given background spacetime~\cite{anderson2005matched,Pound:2009sm,pound2012field,pound2015motion}. In turn, self-force 
theory finds its astrophysical realm in the description of asymmetric 
compact binaries, such as extreme- and intermediate-mass-ratio 
inspirals (EMRIs and IMRIs), which consist of a stellar-mass compact 
object of mass $\mu\sim\ell$ orbiting a more massive primary of mass 
$M\sim{\cal L}$. Their dynamics are dictated by the mass ratio 
$\epsilon=\mu/M\ll 1$, which sets the radiation-reaction timescale 
$t_{\rm RR}=\mathcal{O}(\epsilon^{-1})$ over which the binary 
parameters evolve due to gravitational-wave~(GW) emission.

Asymmetric binaries have become the focus of deep investigation, 
owing to a set of intrinsic properties that make their scientific 
potential particularly appealing. As one of the main observational 
targets of the LISA satellite~\cite{amaroseoane2017laserinterferometerspaceantenna,LISA:2024hlh}, these systems are long-lived, 
and may remain in the sensitivity band of the detector for months to 
years~\cite{LISAConsortiumWaveformWorkingGroup:2023arg}. While in band, they can accumulate $\mathcal{O}(\epsilon^{-1})$ 
orbital cycles on complex trajectories, with large eccentricities and 
inclinations. The richness of the emitted signals, augmented by the 
enormous number of cycles, is expected to enable measurements 
of the source parameters with sub-percent precision~\cite{Barack:2003fp,Babak:2017tow,Burke:2023lno}.

This opens the possibility of using observations of asymmetric 
binaries to probe even the tiniest deviations from GR, which 
can nonetheless pile up over the course of the long orbital evolution. 
Modeling beyond-GR effects requires extending current self-force 
methods to include new fundamental degrees of freedom, possibly coupled 
to the GR ones. The complexity of such calculations has so far 
restricted the development of tractable models to those endowed with 
intrinsic properties that allow drastic simplification.

The most notable example builds upon the inclusion of scalar fields 
non-minimally coupled to gravity, modeled within an effective field 
theory~(EFT) framework~\cite{Maselli:2020zgv,Maselli:2021men,Barsanti:2022vvl,DellaRocca:2024pnm,Gliorio:2026yvh,Spiers:2023cva,Roy:2025kra,barsanti2022extreme,Barsanti:2026ulr,Speri:2024qak,Lestingi:2023ovn}. In this approach, the decoupling of 
scales induced by the large mass asymmetry of the binary simplifies 
the treatment of its dynamics within a genuine self-force approach. 
At leading order in the mass ratio, deviations from the Kerr geometry 
of the primary can be neglected, with the scalar charge of the 
secondary controlling departures from GR~\cite{Maselli:2020zgv}. Notably, the 
effects of the scalar charge can be incorporated as a modular addition 
atop a GR baseline model~\cite{Spiers:2023cva,Barsanti:2026ulr}.

The development of such models at leading~(0PA) order in the mass 
ratio has now reached maturity, with waveforms on generic orbits 
available for LISA parameter estimation~\cite{Gliorio:2026yvh}. However, as for their 
GR counterparts, a 0PA treatment alone is insufficient to avoid 
systematic biases~\cite{Burke:2023lno} in the inference of source parameters from the 
long, cycle-rich signals that characterize these systems,\footnote{A 
fractional error of order $\mathcal{O}(\epsilon)$ in the instantaneous 
frequency, accumulated over $\mathcal{O}(\epsilon^{-1})$ cycles, 
produces an order-unity dephasing, which represents the 
customary accuracy threshold for parameter estimation.} demanding 
that waveform models be pushed to first post-adiabatic~(1PA) order~\cite{Hinderer:2008dm}, 
built within a two-timescale description of the orbital evolution~\cite{Kevorkian1996,Miller:2020bft,pound2022black}.

Such 1PA corrections incorporate first-order (in the mass-ratio) 
conservative and second-order dissipative contributions.%
\footnote{1PA corrections also include the leading terms in the 
evolution of the spin and mass of the primary~\cite{Miller:2020bft}.} 
Moreover, they encode the multipolar structure of the secondary 
beyond its mass monopole --- in particular its linear-in-spin 
contributions stemming from the spin tensor $S^{ab}$~\cite{Mathisson2010,Dixon:1974xoz,papapetrou1951spinning,dixon2015new,Mathews:2021rod,Mathews:2025txc}. 
Beyond their role in waveform accuracy, these multipole moments 
carry direct physical information about the secondary's internal 
structure and, in theories that extend GR, about its coupling to 
additional fundamental degrees of freedom. 
The relevance of these corrections, furthermore, is not confined 
to the EMRI regime: recent results of the self-force program 
have demonstrated its effectiveness in describing 
comparable-mass binaries~\cite{Wardell:2021fyy,Albertini:2022dmc,Mathews:2025txc}, suggesting 
that 1PA waveforms could capture signals from systems targeted 
by future deciHz and ground-based detectors~\cite{ET:2025xjr}.

A comprehensive research effort is currently underway to produce 
1PA waveforms in GR\cite{rosenthal2006second,gralla2012,Pound:2013iba,Pound:2015wva,Miller:2016hjv,Pound:2017psq,vandeMeent:2017bcc,Pound:2019lzj,pound2022black,Upton:2021oxf,Warburton:2021kwk,Wardell:2021fyy,Mathews:2021rod,Spiers:2023cva,Bourg:2024cgh,Cunningham:2024dog,Mathews:2025txc,Upton:2025bja,Spiers:2026yqx,Leather:2026zhl,Musaeus:2026kgp,grant2025flux,Cownden:2026msq,Leather:2026gnc} --- so far limited to quasi-circular orbits 
around slowly spinning primaries~\cite{Wardell:2021fyy,Mathews:2025txc} --- with data-analysis 
studies already highlighting their importance for LISA parameter 
estimation~\cite{Burke:2023lno}. There are multiple complementary programs studying the motion of spinning bodies in GR, including Refs.~\cite{Drummond:2022efc,Drummond:2022xej,Skoupy:2023lih,Drummond:2023loz,Drummond:2026haw,Skoupy:2021asz,Skoupy:2022adh,Skoupy:2024uan,Witzany:2026eqc,piovano2020extreme}.

Additionally, there is progress in extending self-force waveforms from the inspiral phase all the way to merger~\cite{Kuchler:2024esj,Kuchler:2025hwx,Honet:2025dho}, including in scalar-tensor theories of gravity~\cite{Roy:2025kra}.

The scalar-field EFT framework developed in Refs.~\cite{Maselli:2020zgv, Spiers:2023cva} rests on a 
point-particle description of the secondary in which all GR and 
scalar multipoles are encoded by a mass function 
$m(\varphi)$ as the particle Lagrangian. This mass function, 
which carries a direct imprint of the local scalar field along the 
secondary's worldline, suffices to capture the scalar monopole and all effects on the 0PA dynamics of the 
binary; it does not, however, account for any higher multipole 
moment. In particular, it contains neither spin contributions nor 
the richer scalar multipole structure that a spinning secondary 
generically has. An attempt to include GR spin contributions at the level of the equations of motion has been made in \cite{guo2024detecting}, while \cite{Lestingi:2023ovn} included a scalar dipole vector by modeling it as two close scalar monopoles. Here we will pursue a systematic action-based approach and generalize the particle action to include all multipoles that can contribute to 1PA dynamics. This will allow us to derive modified Mathisson-Papapetrou-Dixon (MPD) equations and scalar and gravitational field equations including all terms contributing up to 1PA and hence complete the 1PA treatment 
for spinning bodies endowed with a scalar configuration.

To perform such a derivation  demands, crucially, a systematic methodology to skeletonize 
compact objects dressed with massless bosonic 
fields. This can be done by building on the EFT methods~\cite{Goldberger:2004jt} used to model the motion of extended bodies in GR \cite{Mathisson2010,PhysRevD.73.104031,levi2015spinning}, as first shown in \cite{Loutrel:2018ydv} for dynamical Chern-Simons (dCS) gravity and then for more general scalar-tensor theories in \cite{Brax:2021qqo}.
The starting point is the effective action for a spinning secondary in GR, derived in \cite{PhysRevD.73.104031}, dressed by a scalar field. This approach is a generalization of Eardley's idea to promote the mass of a body appearing in the point particle action to be scalar dependent \cite{Eardley:1975fgi}, capturing the breakdown of the strong equivalence principle. This construction allows us to introduce the body's scalar multipole moments in both the field equations for the metric and the scalar, and in the body's equations of motion, which take the form of modified MPD equations.

While Refs.~\cite{Loutrel:2018ydv,yagi2012post,Brax:2021qqo,almeida2024post} focus on Post-Newtonian studies of dCS and bGR, respectively, we apply this framework to EMRI dynamics. For EMRIs, the scalar charge carried by the primary BH is negligible, as it is suppressed by the primary mass squared \cite{Maselli:2020zgv}. The background scalar field can therefore be taken to be zero, and its leading-order amplitude is governed by the scalar charge of the secondary, and thus proportional to the small mass ratio $\epsilon$ \cite{Maselli:2020zgv}. The scalar dipole contributes at next-to-leading order \cite{Lestingi:2023ovn}.

We find it useful to revisit the derivation of the modified MPD equations and field equations by explicitly defining an effective charge and dipole vector tied to the effective Lagrangian. In doing so, we can exploit the symmetries of the effective Lagrangian to study the structure the effective charge and dipole vector must obey. This lets us decompose them into intrinsic monopole and dipole moments, plus induced corrections in the EMRI limit. In such a symmetry-based analysis, the parity-even scalar and the parity-odd pseudoscalar cases naturally arise as distinct subcases. We find that in the former case, the effective charge affects the body's dynamics up to 1PA only through the intrinsic scalar monopole and a self-coupling correction. The effective dipole enters only at 2PA and can therefore be neglected. The main result is that, up to 1PA, the dynamics depends only on the mass and spin of the secondary, its charge, and the self-coupling constant. In the pseudoscalar case, however, neither the effective charge nor the effective dipole appears at 1PA, and the dynamics is identical to GR.

Throughout this paper we use geometric units, $c=G=1$, 
and the mostly plus signature.

\section{Review of effective action for a spinning body in GR}\label{sec: effective action - spin}

To construct a skeletonized formalism for a spinning body 
endowed with a scalar field, we extend the approach 
developed in Ref.~\cite{PhysRevD.73.104031} which, 
building on a seminal work by Hanson and Regge~\cite{Hanson:1974qy}, 
constructs an effective action for a spinning body in 
GR that leads to the MPD equations upon variation. 

For clarity, we briefly summarize the building 
blocks of the original framework. As a first step, one 
endows the point particle with a spinning frame based on 
tetrads, introducing extra degrees of freedom that describe 
local boosts and rotations. One then defines an antisymmetric 
tensor $\Omega_{\mu \nu}$, interpreted as the body's angular 
velocity, via:
\begin{equation}
    \label{eq: angular velocity and tetrad}
    \Omega_{\mu \nu} = e^{I}_{\mu}\frac{\mathrm{D}}{\mathrm{d} \tau}e_{\nu I}\ .
\end{equation}
Here, $e_{\mu}^I$ is the tetrad frame, with $I \in \{0,1,2,3\}$ 
a local flat index, such that $g_{\mu\nu} e^{\mu}_I e^{\nu}_J = \eta_{IJ}$, 
where $\eta_{IJ}$ is the Minkowski metric. Moreover, 
$\frac{D}{d\tau}:=u^\nu\nabla_\nu$, with $\tau$ a worldline 
parameter which we fix to be the proper time of the 
secondary only at the level of the equations of motion. 
By construction, $\Omega_{\mu\nu}$ is antisymmetric 
due to the orthogonality of the tetrad legs. 

The four-velocity $u^{\mu}$, the angular velocity 
$\Omega_{\mu \nu}$ and the metric $g_{\mu\nu}$ constitute the primitive objects describing the secondary at this order. It can be shown that including the derivative of the metric is only needed to include the mass quadrupole \cite{PhysRevD.73.104031}. 

The core idea of the approach developed in 
\cite{PhysRevD.73.104031} is to use $u$ and $\Omega$ to construct the particle Lagrangian with a 
minimal set of scalars $a_i$ built out of contractions of the two 
cinematic quantities.

Exploiting the Cayley-Hamilton (CH) theorem, 
one can show that $\{a_i\}$ exists and is given by
\begin{equation}
    a_1 = u^2, \quad a_2 = -\Tr(\Omega^2), \quad a_3 = u\Omega^2 u, \quad a_4 = \Tr(\Omega^4)\ ,
\end{equation}
such that any scalar can be written as a function of 
the latter.

The most general Lagrangian for a spinning particle, 
depending on $u$, $\Omega$, and the metric, can 
then be written as a function of $a_i$ as: 
\begin{equation}
    \label{eq: action for spinning particle - generic function of a1,2,3,4}
    S_p = \int \mathrm{d}\tau L(a_1,a_2,a_3,a_4)\ .
\end{equation}
The functional form of $L$ can be further constrained by 
requiring that it is a parity-even scalar and reparameterization 
invariant (RPI). The latter forces the Lagrangian to transform 
under $\tau \rightarrow \tau'$ in such a way that 
$\mathrm{d}\tau' (L(a_1,a_2,a_3,a_4))' = \mathrm{d}\tau L(a_1,a_2,a_3,a_4)$. 
Defining $r = \mathrm{d}\tau/\mathrm{d}\tau'$, RPI then 
imposes that 
\begin{equation}
    L(r^2 a_1, r^2 a_2, r^4 a_3, r^4 a_4) = r L(a_1,a_2,a_3,a_4)\ .
\end{equation}
By varying the action with respect to the coordinates 
and the tetrad, $\delta S_p/\delta x^{\mu}$ and $e^{\nu I}\delta S_p/\delta e^{I}_{\mu}$, 
Ref.~\cite{PhysRevD.73.104031} derives the MPD 
equations of motion without the need to specify 
the form of the Lagrangian:
\begin{align}
    \label{eq: MPD equations - linear momentum equation}
    \frac{\mathrm{D}p^{\mu}}{\mathrm{d}\tau} &= - \frac{1}{2}R^{\mu \alpha \beta \gamma} S_{\beta \gamma}u_{\alpha} ,
     \\ \label{eq: MPD equations - spin tensor equation} 
    \frac{\mathrm{D}S^{\mu \nu}}{\mathrm{d}\tau} &= 2 p^{[\mu}u^{\nu]}\ ,
\end{align}
where $p^{\mu}$ and $S_{\mu \nu}$ are defined in terms of $u$ and $\Omega$ as:
\begin{align}
    \label{eq: Porto's linear momentum}
p^{\mu} =& 2 u^{\mu} \frac{\partial L}{\partial a_1} + 2 \Omega^{\mu \nu}\Omega_{\nu \alpha}u^{\alpha}
\frac{\partial L}{\partial a_3}\ ,
\\  \label{eq: Porto's spin tensor}
S^{\mu \nu} =& 4 \Omega^{\mu \nu} \frac{\partial L}{\partial a_2} +2(u^{\mu}\Omega^{\nu \alpha}u_{\alpha}-u^{\nu}\Omega^{\mu \alpha}u_{\alpha})\frac{\partial L}{\partial a_3} 
\nonumber \\ &
+8 \Omega^{\nu \beta} \Omega_{\beta \alpha}\Omega^{\alpha \mu}\frac{\partial L}{\partial a_4}\ .
\end{align}
Note that to cast \eqref{eq: MPD equations - spin tensor equation} in 
this form, one exploits the identity:
\begin{equation}
\label{eq: magic antisymmetrization formula}
    S^{[\mu|\alpha|}\Omega_{\alpha}^{\ \ \nu]} - p^{[\mu}u^{\nu]}=0\ .
\end{equation}
It is also useful to note that $\Tr(S^2)$ is conserved on the 
worldline, i.e.\ 
\begin{equation}
\label{eq: conservation of S^2}
    \frac{\mathrm{D}(\Tr(S^2))}{\mathrm{d}\tau}= 0\ .
\end{equation}
At this stage, the system \eqref{eq: MPD equations - linear momentum equation}-\eqref{eq: MPD equations - spin tensor equation} features $10$ equations for $13$ unknowns, and is hence underdetermined. 
The tetrad indeed introduces $6$ extra degrees of freedom, corresponding 
to local boosts and rotations. The system can be closed by imposing 
the so-called \textit{spin supplementary condition} (SSC), which removes 
these extra degrees of freedom~\cite{PhysRevD.73.104031,Barausse:2009aa}:
\begin{equation}
    \label{eq: SSC}S^{\mu\nu}p_{\nu}=0\ .
\end{equation}
Moreover, for this constraint to be preserved under the evolution of the system, 
the following relation must hold:
\begin{equation}
\label{eq: linear momentum - spin square relation}
  (p^{\nu}u_{\nu})  p^{\mu} = (p^{\nu}p_{\nu}) u^{\mu} + \frac{1}{2}R_{\nu \lambda \rho \sigma}S^{\mu \nu}S^{\rho \sigma}u^{\lambda}\ .
\end{equation}
Upon variation of the action \eqref{eq: action for spinning particle - generic function of a1,2,3,4} 
with respect to the metric, and defining 
$T_{\mu\nu} = -\frac{16 \pi}{\sqrt{-g}}\frac{\delta S_p}{\delta g^{\mu\nu}}$, 
we recover the well-known pole-dipole approximation for the particle 
stress-energy tensor:
\begin{align}
\label{eq: pole-dipole stress-energy tensor}
    T_{\mu \nu} =& 8 \pi \int \mathrm{d}\tau \frac{\delta^{(4)}(x-z(\tau))}{\sqrt{-g}} p_{( \mu}u_{\nu )}
    \nonumber \\ &
    - 8 \pi \nabla^{\rho} \int \mathrm{d}\tau \frac{\delta^{(4)}(x-z(\tau))}{\sqrt{-g}} S_{\rho (\mu}u_{\nu )}\ .
\end{align}
Equation~\eqref{eq: pole-dipole stress-energy tensor} provides 
precisely the leading terms in the multipolar expansion of
the stress-energy tensor of an extended secondary, 
as originally devised by Tulczyjew~\cite{Tulczyew}:
\begin{align}
\label{eq: Tulczyjew - multipole expansion of stress-energy tensor}
    T^{\mu \nu} =&  \int \mathrm{d}\tau \frac{\delta^{(4)}(x-z(\tau))}{\sqrt{-g}} t^{\mu \nu} 
    \nonumber \\ &
+ \nabla_{\alpha_1}\int \mathrm{d}\tau \frac{\delta^{(4)}(x-z(\tau))}{\sqrt{-g}} t^{\mu \nu \alpha_1} + ...
\end{align}

As shown in e.g.~\cite{Steinhoff:2009tk}, the conservation 
law $\nabla_{\mu} T^{\mu \nu} = 0$ allows one to derive the 
equations of motion for the pole $t^{\mu \nu}$ and the 
dipole $t^{\mu \nu \alpha_1}$, together with the associated 
constraint equations.

\section{Effective action for a spinning body with scalar multipoles}\label{sec: effective action - multipoles}

We now introduce the relevant formalism to extend previous 
results to a secondary with scalar multipole moments. 

We follow the EFT construction of \cite{Maselli:2020zgv,Spiers:2023cva}, 
considering the action for a real, massless, scalar field 
non-minimally coupled to gravity:
\begin{equation}
\label{eq: general action}
    S[g,\varphi,\Psi] = S_0[g,\varphi] + \alpha S_c[g,\varphi]+S_m[g,\varphi,\Psi]\ .
\end{equation}
where $\Psi$ accounts for possible matter fields. The action
$S_c$ introduces the non-minimal coupling, with the parameter
$\alpha$ being dimensionful such that in geometric units
$[\alpha]=[mass^n]$, $n \in \mathbb{N}$, $n \geq 2$. 
Moreover, in the limit $\alpha \rightarrow 0$, BH solutions are continuously connected to Kerr in 
GR (we refer the reader to
\cite{Spiers:2023cva} for further technical details).
Finally, $S_0$ is the Einstein-Hilbert action,  
also containing the scalar field kinetic term:
\begin{equation}
    S_0[g,\varphi] = \frac{1}{16 \pi} \int \mathrm{d}^4 x \sqrt{-g} (R - \frac{1}{2}\nabla_{\mu}\varphi \nabla^{\mu}\varphi)\ .
\end{equation}

We then specialize to the case where $S_m[g,\varphi,\Psi] \equiv S_p[g,\varphi]$ describes a spinning object endowed with a scalar multipolar structure. The construction of $S_p[g,\varphi]$ in this scenario is the central matter of this section.

After variation of the action, the scalar field 
obeys the field equation:
\begin{equation}
\label{eq: scalar field equations - generic}
    \Box \varphi = \alpha \Sigma_c + \Sigma\ ,
\end{equation}
where 
\begin{equation}
\label{eq: defining scalar source from action variation}
    \Sigma = -\frac{16 \pi}{\sqrt{-g}}\frac{\delta S_p}{\delta \varphi}, \quad \Sigma_c = -\frac{16 \pi}{\sqrt{-g}}\frac{\delta S_c}{\delta \varphi}\ .
\end{equation}

We now extend Tulczyjew's approach from the 
stress-energy tensor to the scalar sector, introducing a 
multipolar expansion of the scalar field distribution $\Sigma$. 
In particular, we introduce an \textit{effective} scalar monopole $Q$ and  scalar dipole vector $K^{\mu}$
such that
:
\begin{align}
\label{eq: Tulczyjew-like multipolar expansion of scalar-charge distribution}
    \Sigma =& -16 \pi \int \mathrm{d}\tau Q \frac{\delta^{(4)}(x^{\alpha}-z^{\alpha}(\tau))}{\sqrt{-g}} +
    \nonumber \\ &
    + 16 \pi \nabla_{\mu} \int \mathrm{d}\tau K^{\mu} \frac{\delta^{(4)}(x^{\alpha}-z^{\alpha}(\tau))}{\sqrt{-g}} + ... \ .
\end{align}

Unlike the pole-dipole terms of the gravitational stress-energy tensor, 
$\Sigma$ obeys no conservation law from which equations of motion for 
the multipoles $Q, K, \ldots$ can be derived. Consequently, the scalar 
multipoles are not determined by the dynamics alone. 
They are system-dependent quantities whose values must be 
prescribed, encoding the internal physics of the body. We also note how our definition of the scalar multipoles via \eqref{eq: Tulczyjew-like multipolar expansion of scalar-charge distribution} is such that the charge contains all $\ell=0$ multipoles, $K^{\mu}$ all $\ell=1$ multipoles, and so forth. 

In the remainder of this section, we describe 
how to construct an effective action that captures the multipolar expansion of the scalar-charge distribution \eqref{eq: Tulczyjew-like multipolar expansion of scalar-charge distribution}. 
In doing so, we directly relate the effective multipoles $Q,K,...$ to the effective Lagrangian describing the extended object. 
The main goal is to generalize the approach reviewed in Sec.~\ref{sec: effective action - spin}. However, we will not attempt to introduce new tensors that relate to the scalar multipoles directly in the action (such as $\Omega^{
\mu\nu}$ for the spin in GR), but instead we will ``dress up'' action \eqref{eq: action for spinning particle - generic function of a1,2,3,4} with the scalar field, which we will continue to treat as a dynamical variable alongside the metric. 

This will result in an extension of Eardley's effective action for a point particle in scalar-tensor theories of gravity \cite{Eardley:1975fgi} to a spinning secondary. There, Eardley implements the breakdown of the strong equivalence principle by dressing the particle action with the scalar field by promoting $m \rightarrow m(\varphi)$. Once we have put together our action, we will derive the equations of motion for a spinning object with a scalar pole-dipole structure. This procedure generalizes the MPD equations, showing how the dynamics of the body is affected by the scalar field. 

We note that a similar derivation, also inspired by 
the EFT techniques, was used in Ref.~\cite{Loutrel:2018ydv} 
to model the motion of compact objects, although in 
the specific case of dCS gravity. In Sec.~\ref{subsec: generalized MPD equations} we show that our 
formalism, while general, reproduces the modified MPD equations derived in \cite{Loutrel:2018ydv} in the 
relevant limit.

\subsection{Effective action for a nonspinning body with a scalar charge}\label{sec: effective action - monopole}

We start by constructing an effective action $S_p$ for a non-spinning body 
dressed by a scalar field exploiting the formalism described in 
Sec.~\ref{sec: effective action - spin}. We then show that this formulation 
matches the original effective action first advocated by Eardley \cite{Eardley:1975fgi} and adopted in \cite{Maselli:2020zgv,Spiers:2023cva}. 
Rather than the angular velocity tensor $\Omega$, 
we take the scalar field $\varphi$ as one of the primitive 
quantities defined on the worldline of the secondary from which to build 
the minimal set of scalars needed for $S_p$. In particular, 
defining 
\begin{equation}
    a_1 = u^2, \quad b_1 = \varphi\ ,
\end{equation}
the effective action becomes
\begin{equation}
    S_p = \int \mathrm{d}\tau L(u^2,\varphi) \ .
\end{equation}
Imposing RPI, the latter reduces to:
\begin{equation}
    \label{eq: effective action - monopole}
    S_p=\int \mathrm{d}\tau\, m(\varphi)\sqrt{-u^2}\ ,
\end{equation}
where $m(\varphi)$ is a generic function of the 
scalar field, required to reduce to minus the bare mass of the object
$\mu$ in the GR 
limit $\varphi \rightarrow 0$. This action is exactly the one first introduced by Eardley by promoting the mass of the object to be scalar field dependent $m \rightarrow -m(\varphi)$ \cite{Eardley:1975fgi}. 

Plugging $S_p$ in \eqref{eq: defining scalar source from action variation}, 
we find for the scalar charge distribution $\Sigma$:
\begin{align}
\label{eq: scalar charge distribution - nonspinning particle with charge}
    \Sigma = -16 \pi \int \mathrm{d}\tau \frac{\delta^{(4)}(x^{\alpha}-z^{\alpha}(\tau))}{\sqrt{-g}}\frac{\partial m(\varphi)}{\partial \varphi}\ .
\end{align}
Matching this expression with the leading term of the 
multipolar decomposition \eqref{eq: Tulczyjew-like multipolar expansion of scalar-charge distribution}, we find:
\begin{equation}
    Q=\frac{\partial L}{\partial \varphi}=\frac{\partial m}{\partial \varphi}\ .
\end{equation}
We can thus interpret $\partial m(\varphi)/\partial \varphi$ 
as the effective scalar charge of the body. At this stage, 
the charge is a free parameter of the model, 
depending on the function $m(\varphi)$, 
which encodes the specific non-minimal coupling 
in~\eqref{eq: general action} and is not fixed 
by the equations of motion.

The effective action \eqref{eq: effective action - monopole} 
cannot accommodate the scalar dipole vector $K$ or 
higher-order multipoles, as these require introducing 
a preferred direction for the field, encoded, for instance, 
by $\nabla_{\mu}\varphi$. This is what we 
turn to in the next subsection.

\subsection{Effective action for a spinning body with scalar monopole-dipole structure}\label{sec: effective action - multipoles and spin}

We now extend previous results to a spinning 
body, considering all the scalars that can 
be built using $g,u,\Omega,\varphi,\nabla \varphi$:
\begin{equation}
    \{a_1,a_2,a_3,a_4,b_1,b_2\}\ , \quad b_2=k^{\mu}(g,u,\Omega,\nabla\varphi)\nabla_\mu \varphi\ .
\end{equation}
Our construction relies on $(g,u,\Omega)$ 
providing the structure for the pole-dipole 
gravitational sector, as shown in Ref.~\cite{PhysRevD.73.104031}, 
with $(\varphi, \nabla \varphi)$ yielding 
the scalar pole-dipole distribution. 
We require $k^{\mu}$ to be constructed from $u,\Omega,\nabla\varphi$ in such a way that 
it is a vector.
Then, we write the effective action for a 
spinning body with a scalar monopole and 
dipole vector as:
\begin{equation}
    \label{eq: effective action - spin and pole-dipole}
    S_p=\int \mathrm{d}\tau L(a_i,\varphi,k^{\mu}\nabla_{\mu}\varphi)\ .
\end{equation}
Note that $k^{\mu}$ needs to ensure the RPI of 
the Lagrangian, as discussed before. Moreover 
we assume it is independent of $\varphi$, as 
the Lagrangian $L$ already contains 
$\varphi$ in a general way. 

While the form of the coefficients 
$a_i$ is specified, the structure of 
$k^\mu$ is currently 
left undetermined. 
We will be able to specify it in the EMRI limit (see \cref{subsec:scalar-vs-pseudoscalar}), where the 
CH theorem and a gradient expansion up to 
second order in the mass-ratio constrain the form of 
$k^{\mu}$ and of the Lagrangian to depend only on 
$a_i,\varphi$ and $\nabla_\mu\varphi$ in a prescribed way. 

As we show in App.~\ref{app: Scalar-charge distribution from effective action}, varying 
$S_p$ in Eq.~\eqref{eq: effective action - spin and pole-dipole} with respect to the scalar field yields the pole-dipole scalar-charge distribution
\begin{align}
\label{eq: scalar charge distribution - spinning particle with dipole}
    \Sigma =&  -16 \pi \int \mathrm{d}\tau \frac{\partial L}{\partial \varphi}\frac{\delta^{(4)}(x^{\alpha}-z^{\alpha}(\tau))}{\sqrt{-g}} +
    \nonumber \\ &
    + 16 \pi \nabla_{\mu}\int \mathrm{d}\tau \frac{\partial L}{\partial \nabla_\mu \varphi}\frac{\delta^{(4)}(x^{\alpha}-z^{\alpha}(\tau))}{\sqrt{-g}}\ .
\end{align}
Comparing the latter with the multipolar expansion 
in Eq.~\eqref{eq: Tulczyjew-like multipolar expansion of scalar-charge distribution}, we can identify 
the effective scalar monopole and dipole vector
\begin{equation}
\label{eq: general dipole vector}
   Q(a_i,b_j)=\frac{\partial L}{\partial \varphi}\ , \quad K^{\mu}(g,u,\Omega,\varphi,\nabla \varphi) =\frac{\partial L}{\partial \nabla_{\mu} \varphi}\ .
\end{equation}
As defined, the effective scalar 
dipole vector is not spacelike. Nevertheless, as we show in App.~\ref{app: Scalar-charge distribution from effective action}, the scalar distribution $\Sigma$ 
in Eq.~\eqref{eq: scalar charge distribution - spinning particle with dipole} is invariant under 
the transformation
\begin{align}
    \label{eq: redundancy in scalar pole-dipole distribution}
    Q \rightarrow Q - \frac{\mathrm{d} f}{\mathrm{d}\tau}, \qquad K^{\mu} \rightarrow K^{\mu} - f u^{\mu}.
\end{align}
We can therefore exploit this redundancy to make the dipole vector spacelike, a 
property that needs to be satisfied by any 
physical dipole moment. Indeed, exploiting the decomposition of $K^{\mu}$ with respect to the four-velocity, $K^{\mu} = K_{\parallel}(\tau) u^{\mu} + K^{\mu}_{\perp}$, $K^{\mu}_{\perp} = (\delta^{\mu}_{\nu}+u^{\mu}u_{\nu})K^{\nu}$, and choosing $f(\tau) = K_{\parallel}(\tau) = -K^{\mu}u_{\mu}$, we have
\begin{align}
\label{eq: scalar charge distribution - spinning particle with dipole - physical definition}
    \Sigma =&  -16 \pi \int \mathrm{d}\tau \tilde Q\frac{\delta^{(4)}(x^{\alpha}-z^{\alpha}(\tau))}{\sqrt{-g}} +
    \nonumber \\ &
    + 16 \pi \nabla_{\mu}\int \mathrm{d}\tau K^{\mu}_{\perp}\frac{\delta^{(4)}(x^{\alpha}-z^{\alpha}(\tau))}{\sqrt{-g}}\ ,
\end{align}
where
\begin{align}
\label{eq: physical charge}
     \tilde Q &= \frac{\partial L}{\partial \varphi} + \frac{\mathrm{d}}{\mathrm{d}\tau}\left( \frac{\partial L}{\partial \nabla_\mu \varphi}u_{\mu} \right),
     \\
  \label{eq: physical dipole}
     K^{\mu}_{\perp} &= (\delta^\mu_\nu + u^{\mu}u_\nu)\frac{\partial L}{\partial \nabla_{\nu}\varphi}\ .
\end{align}
While $\tilde{Q}$ and $K^{\mu}_{\perp}$ provide an unambiguous definition of the effective monopole and dipole respectively, we will be using $Q$ and $K^{\mu}$ as well below, with the understanding that one can always switch to $\tilde{Q}$ and $K^{\mu}_{\perp}$ by using eq.~\eqref{eq: redundancy in scalar pole-dipole distribution} and choosing $f(\tau) = K_{\parallel}(\tau)$.

\section{MPD equations in presence of a scalar monopole and dipole}\label{subsec: generalized MPD equations}

We proceed to compute how the MPD equations 
change when the body carries an 
effective scalar charge and a dipole vector. 
As discussed in Sec.~\ref{sec: effective action - spin}, 
we need to determine the variations of the effective 
action with respect to the position and 
tetrad, $\delta_z S$ and $\delta_e S$. 
Explicit computation of $\delta_z L$ yields
\begin{align}
\label{eq: delta-x L}
    \delta_z L  = &\frac{\partial L}{\partial a_i}\delta_z a_i + \frac{\partial L}{\partial b_i}\delta_z b_i 
    \nonumber \\ 
    = &\frac{\partial L}{\partial a_i}\left(\frac{\partial a_i}{\partial g_{\mu \nu}}\delta_z g_{\mu \nu} +  \frac{\partial a_i}{\partial u^{\mu}}\delta_z u^{\mu} + \frac{\partial a_i}{\partial \Omega_{\mu \nu}}\delta_z \Omega_{\mu \nu} \right)
    \nonumber \\ 
    &+\frac{\partial L}{\partial b_2}\left(\frac{\partial b_2}{\partial g_{\mu \nu}}\delta_z g_{\mu \nu} +  \frac{\partial b_2}{\partial u^{\mu}}\delta_z u^{\mu} + \frac{\partial b_2}{\partial \Omega_{\mu \nu}}\delta_z \Omega_{\mu \nu} \right)
    \nonumber \\ 
    &+\frac{\partial L}{\partial b_1}\frac{\partial b_1}{\partial \varphi}\delta_z \varphi + \frac{\partial L}{\partial b_2} \frac{\partial b_2}{\partial \nabla_{\mu}\varphi}\delta_z \nabla_{\mu}\varphi\ .
\end{align}
It is then natural to introduce the generalized linear 
momentum $\mathcal{P}^{\mu}$ and spin tensor 
$\mathcal{S^{\mu \nu}}$ as
\begin{align}
\label{eq: generalized linear momentum - main tex}
    \mathcal{P}^{\mu} &= \frac{\partial L}{\partial a_i}\frac{\partial a_i}{\partial u_\mu}+\frac{\partial L}{\partial b_2}\frac{\partial b_2}{\partial u_\mu}\ ,
    \\
\label{eq: generalized spin tensor - main tex}
    \mathcal{S^{\mu \nu}}&=2\frac{\partial L}{\partial a_i}\frac{\partial a_i}{\partial \Omega_{\mu \nu}}+2\frac{\partial L}{\partial b_2}\frac{\partial b_2}{\partial \Omega_{\mu \nu}}\ .
\end{align}
Plugging these definitions into \eqref{eq: delta-x L} yields
\begin{align}
\label{eq: final delta-x L}
    \delta_z L  =& \mathcal{P}_{\mu}\delta_z u^{\mu} +\frac{1}{2}\mathcal{S}^{\mu \nu}\delta_z \Omega_{\mu \nu} + \frac{\partial L}{\partial g_{\mu \nu}}\delta_z g_{\mu \nu}
    \nonumber \\ &
    + \frac{\partial L}{\partial \varphi} \delta_z \varphi + \frac{\partial L}{\partial \nabla_{\mu}\varphi}\delta_z \nabla_{\mu}\varphi.
\end{align}
Similarly, varying $L$ with respect to the 
tetrad and using 
$\delta_e g= \delta_e u = \delta_e \varphi = \delta_e \nabla_{\mu}\varphi = 0$, 
we have:
\begin{equation}
\label{eq: final delta-e L}
    \delta_e L =  \frac{1}{2} \mathcal{S}^{\mu \nu} \delta_e \Omega_{\mu \nu}\ .
\end{equation}
Equations~\eqref{eq: final delta-x L}-\eqref{eq: final delta-e L}
allow us to determine the variation of the effective action 
and the equations of motion for 
$\mathcal{P}^\mu$ and $\mathcal{S}^{\mu\nu}$ 
(we refer the reader to 
App.~\ref{app: Modified MPD equations from effective action} 
for the details of the derivation):
\begin{align}
\label{eq: generalised MPD equation - linear momentum}
    \frac{\mathrm{D}\mathcal{P}^{\mu}}{\mathrm{d}\tau} &= -\frac{1}{2}R^{\mu \alpha \beta \gamma}\mathcal{S}_{\beta \gamma}u_{\alpha} + Q\nabla^{\mu}\varphi + K^{\rho}\nabla_{\rho}\nabla^{\mu}\varphi,
     \\
     \label{eq: generalised MPD equation - spin tensor}
    \frac{\mathrm{D}\mathcal{S}^{\mu \nu}}{\mathrm{d}\tau} & = 2 \mathcal{S}^{[\mu| \alpha|}\Omega_{ \alpha}^{\ \ \nu]} =2 \mathcal{P}^{[\mu}u^{\nu]}+2 K^{[\mu}\nabla^{\nu ]}\varphi,
\end{align}
where the last equality in 
Eq.~\eqref{eq: generalised MPD equation - spin tensor} 
generalises Eq.~\eqref{eq: magic antisymmetrization formula} and is the antisymmetrization of the identity \eqref{eq:magic-relation} derived in App.~\ref{app: Dipole correction to spin-linear momentum relations}. 

We proceed to highlight the main features of the 
modified MPD equations. First, they reduce to the 
standard equations \eqref{eq: MPD equations - linear momentum equation}-\eqref{eq: MPD equations - spin tensor equation} 
as $\varphi \rightarrow 0$, for which 
$\mathcal{S}_{\mu \nu}\rightarrow S_{\mu \nu}$ and $\mathcal{P}^{\mu}\rightarrow p^{\mu}$.
Then, while Eqns.~\eqref{eq: generalised MPD equation - linear momentum}-\eqref{eq: generalised MPD equation - spin tensor} retain 
the same structure as the standard MPD 
equations - with $p^{\mu},S_{\mu \nu}$ replaced 
by the generalised linear momentum vector and 
spin tensor - they feature novel terms 
that depend on the effective charge 
$Q$ and on the scalar dipole vector 
$K^{\mu}$. We interpret the terms 
$Q \nabla^{\mu}\varphi$ and $K^{\rho}\nabla_{\rho}\nabla^{\mu}\varphi$ in Eq.~\eqref{eq: generalised MPD equation - linear momentum} 
as scalar self-force contributions, 
proportional to the scalar monopole and dipole, 
respectively. Furthermore, we find that the precession of $\mathcal{S}_{\mu \nu}$ 
is driven by the misalignment between 
$\mathcal{P}^{\mu}$ and $u^{\mu}$, as in GR, 
and by the misalignment between the dipole vector 
of the object and the scalar field gradient. 

It is easy to see that $\mathcal{S}^{\mu\nu}\mathcal{S}_{\mu\nu}$ 
is conserved along the worldline
\begin{align}
    \frac{\mathrm{D}(\mathcal{S}^{\mu\nu}\mathcal{S}_{\mu\nu})}{\mathrm{d\tau}} &= 2(\mathcal{S}^{\mu \lambda}\Omega_{\lambda}^{\ \ \nu}-\Omega^{\mu}_{\ \lambda}\mathcal{S}^{\lambda \nu})\mathcal{S}_{\mu \nu}
    \nonumber \\ &
    =4 \Tr(\Omega \mathcal{S}^2)=0\ ,
\end{align}
where in the last step we exploited the antisymmetry 
of $\Omega$ and the symmetry of $\mathcal{S}^2$. 
Hence, on the worldline
\begin{equation}
\label{eq: conservation of generalised S^2}
    \mathcal{S}^{\mu\nu}\mathcal{S}_{\mu\nu} = c_1\ ,
\end{equation}
for some constant $c_1$.

Following the discussion of Sec.~\ref{sec: effective action - spin}, 
one immediately sees that the modified MPD 
equations \eqref{eq: generalised MPD equation - linear momentum}, \eqref{eq: generalised MPD equation - spin tensor} also form an 
undetermined system, as in GR. While $\varphi$ can 
be determined from its field equation, we still 
have a system of $10$ equations for $13$ unknowns. 
Analogously to GR, it is hence natural to impose 
the following generalized SSC:
\begin{equation}
    \label{eq: generalised SSC}
    \mathcal{S}^{\mu \nu}\mathcal{P}_{\nu} = 0,
\end{equation}
which reduces to the standard SSC condition, 
Eq.~\eqref{eq: SSC}, as $\varphi \rightarrow 0$.
Requiring that the generalized SSC \eqref{eq: generalised SSC} 
is preserved as the system evolves leads to the 
following relation:
\begin{align}
    \label{eq: modified linear momentum - post generalised SSC}
    (\mathcal{P}^{\nu}u_{\nu})\mathcal{P^{\mu}} =& (\mathcal{P}^{\nu}\mathcal{P}_{\nu})u^{\mu} - \mathcal{S}^{\mu\nu}\frac{\mathrm{D}\mathcal{P}_{\nu}}{\mathrm{d}\tau} 
    \nonumber \\ &
    + (K^{\nu}\mathcal{P}_{\nu})\nabla^{\mu}\varphi - (\nabla^{\nu}\varphi \mathcal{P}_{\nu})K^{\mu}\ .
\end{align}

In the case of a nonspinning particle and 
neglecting the $\nabla\varphi$ contributions, 
we showed in Sec.~\ref{sec: effective action - monopole} 
that $L=m(\varphi)\sqrt{-u^2}$.
In this case, Eqs.~\eqref{eq: generalized linear momentum - main tex} and 
\eqref{eq: generalised MPD equation - linear momentum}
yield:
\begin{equation}
\label{eq: generalized linear momentum - only scalar charge and no spin}
    \mathcal{P}^{\mu} = - \frac{m(\varphi)}{\sqrt{-u^2}}u^{\mu}\ ,
\end{equation}
and
\begin{equation}
    \frac{\mathrm{D}\mathcal{P}^{\mu}}{\mathrm{d}\tau} = Q\nabla^{\mu}\varphi = \frac{\partial m}{\partial \varphi}\nabla^{\mu}\varphi\ .
\end{equation}
Finally, using \eqref{eq: generalized linear momentum - only scalar charge and no spin}, we get
\begin{equation}
    m(\varphi)\frac{\mathrm{D}u^{\mu}}{\mathrm{d}\tau} = -(g^{\mu \nu} + u^{\mu}u^{\nu})\frac{\partial m}{\partial \varphi} \nabla_\nu \varphi\ ,
\end{equation}
which recovers the result of \cite{quinn2000axiomatic,Zimmerman:2015rga,Spiers:2023cva}, 
describing the scalar self-force term proportional to the charge, 
which shifts the body away from its geodesic motion.

As another important check, we note that setting $Q=0$ and $K^{\rho}\sim S^{\rho}$ 
the set of equations \eqref{eq: generalised MPD equation - linear momentum}-\eqref{eq: generalised MPD equation - spin tensor}
reduces to the modified MPD equations derived in the context 
of dCS gravity in Ref.~\cite{Loutrel:2018ydv}.
We also note that \cref{eq: generalised MPD equation - linear momentum,eq: generalised MPD equation - spin tensor} have a similar structure to the modified MPD equations in Ref.~\cite{obukhov2014equations} (Eqns.~(32)-(33)), derived for even-parity scalar-tensor theories from stress-energy tensor balance laws.

\section{Scalar pole-dipole-induced corrections to the Einstein equation}\label{subsec: generalized stress-energy tensor}

Consider the action \eqref{eq: general action}. Then, the Einstein equation reads 
\begin{equation}
    \label{eq: Einstein equation for general action}
    G_{\mu\nu}[g] = T_{\mu\nu}^{\text{scal}}[g,\varphi] + \alpha T_{\mu\nu}^{c}[g,\varphi]+T_{\mu\nu}^p[g,\varphi], 
\end{equation}
where
\begin{equation}
    T_{\mu\nu}^{\text{scal}} = \frac{1}{2}\nabla_{\mu}\varphi \nabla_{\nu}\varphi - \frac{1}{4} g_{\mu\nu}\nabla^{\rho}\varphi\nabla_{\rho}\varphi,
\end{equation}
and 
\begin{equation}
\label{eq:stressenergycp}
    T_{\mu\nu}^{c,p}= -\frac{16\pi}{\sqrt{-g}} \frac{\delta S_{c,p}}{\delta g^{\mu\nu}}.
\end{equation}
Consider the effective action $S_p = \int \mathrm{d}\tau L(a_i,\varphi,k^{\mu}\nabla_{\mu}\varphi)$. We compute the stress-energy tensor $T_{\mu \nu}^p$ by varying the action with respect to the inverse metric. Using
\begin{align}
    \delta_g L =& \frac{\partial L}{\partial a_i}\left( \left . \frac{\partial a_i}{\partial g^{\mu \nu}}\right |_\Omega \delta g^{\mu \nu} + \frac{\partial a_i}{\partial \Omega_{\beta \alpha}}\delta_g \Omega_{\beta \alpha} \right) \nonumber \\ &
    +\frac{\partial L}{\partial b_2} \left( \left.\frac{\partial b_2}{\partial g^{\mu \nu}}\right|_{\Omega} \delta g^{\mu \nu} + \frac{\partial b_2}{\partial \Omega_{\beta \alpha}}\delta_g \Omega_{\beta \alpha} \right)
    \\
    \label{eq:variation-angular-velocity}
    \delta_g \Omega_{\beta \alpha} =& u^{\lambda}\nabla_{[\beta}\delta g_{\alpha ] \lambda} + g_{\mu [ \beta}\Omega_{\alpha ] \nu} \delta g^{\mu \nu} \ ,
\end{align}
where $|_{\Omega}$ means that $\Omega$ is being held fixed with respect to the metric,
we find (see App.~\ref{app: Deriving pole-dipole stress-energy tensor from effective action})
\begin{align}
\label{eq:stressenergy-beforemagicformula}
    T_{\mu \nu}^p =& -16 \pi \int \mathrm{d}\tau \frac{\delta^{(4)}(x-z(\tau))}{\sqrt{-g}} \left( \left .\frac{\partial L}{\partial g^{\mu \nu}} \right
    |_\Omega  \right.
    \nonumber \\ &
    \left. + \frac{1}{2}\mathcal{S}_{(\mu | \alpha}\Omega^{\alpha}_{\ \ \nu)}\right) 
    - 8 \pi \nabla^{\rho} \int \mathrm{d}\tau \frac{\delta^{(4)}(x-z(\tau))}{\sqrt{-g}} \mathcal{S}_{\rho (\mu}u_{\nu )},
\end{align}
where 
\begin{equation}
\label{eq:lagrangian-metric-derivative}
    \left. \frac{\partial L}{\partial g^{\mu \nu}} \right
    |_\Omega =   \frac{\partial L}{\partial a_i} \left . \frac{\partial a_i}{\partial g^{\mu \nu}}\right |_\Omega + \frac{\partial L}{\partial b_2} \left . \frac{\partial b_2}{\partial g^{\mu \nu}}\right |_\Omega .
\end{equation}
Furthermore, in App.~\ref{app: Dipole correction to spin-linear momentum relations} we show that
\begin{equation}
   \left . 2\frac{\partial L}{\partial g^{\mu \nu}}\right |_\Omega + \mathcal{S}_{(\mu | \alpha}\Omega^{\alpha}_{\ \ \nu)} = -\mathcal{P}_{( \mu}u_{\nu )} + K_{(\mu}\nabla_{\nu )}\varphi .
\end{equation}
Using this identity, we finally find the stress-energy tensor:
\begin{align}
\label{eq: stress-energy tensor with scalar dipole correction}
    T_{\mu \nu}^p =& 8 \pi \int \mathrm{d}\tau \frac{\delta^{(4)}(x-z(\tau))}{\sqrt{-g}} \left( \mathcal{P}_{( \mu}u_{\nu )} - K_{(\mu}\nabla_{\nu )}\varphi \right) 
    \nonumber \\ &
    - 8 \pi \nabla^{\rho} \int \mathrm{d}\tau \frac{\delta^{(4)}(x-z(\tau))}{\sqrt{-g}} \mathcal{S}_{\rho (\mu}u_{\nu )}.
\end{align}
Strikingly, $T_{\mu \nu}^p$ has the same structure as the GR pole-dipole stress-energy tensor, \eqref{eq: pole-dipole stress-energy tensor}, with $\{p^{\mu},S_{\mu \nu}\}\rightarrow \{\mathcal{P}^{\mu},\mathcal{S}_{\mu \nu}\}$ but also features the explicit dipole vector correction $K_{(\mu}\nabla_{\nu )}\varphi$.

\section{Scalar vs pseudoscalar}\label{subsec:scalar-vs-pseudoscalar}
The way we relate the multipoles to the effective Lagrangian is the main strength of our new construction, tying the structure of the 
monopole and dipole to minimal sets of scalars and vector quantities.

In what follows we specialize to the following scalar field expansion of the
effective Lagrangian
\begin{equation}
L = L_{\text{GR}}(a_i) + L_{\varphi},
 \end{equation}
 where $L_{\text{GR}}$ has the same structure described in \ref{sec: effective action - spin}, and $L_\varphi$ is 
 \begin{equation}
\label{eq: effective lagrangian with phi - phi^2 - grad phi}
    L_{\varphi} = (Q_1(g,u,\Omega) + \frac{1}{2}Q_2(g,u,\Omega)\varphi)\varphi + K^{\mu}(g,u,\Omega)\nabla_{\mu}\varphi.
\end{equation}
This is the relevant effective Lagrangian describing the secondary in the EMRI setting discussed in Sec.~\ref{sec: EMRI case}. As we will explain later in more detail, in this case
we have a natural expansion parameter, the mass-ratio $\epsilon$, and the scaling $\varphi \sim O(\epsilon)$. As our aim is to derive the dynamics of the secondary including all terms up to $O(\epsilon^2)$, we need to retain all such terms in the Lagrangian as well, but can neglect higher order terms. The scaling of the scalar field with the mass ratio immediately implies that we only need the terms linear and quadratic in the field. Now consider the term $K^{\mu}(g,u,\Omega)\nabla_{\mu}\varphi$. The gradient of the field scales as $\epsilon/M$, with $M$ the lengthscale of the primary. Dimensional analysis implies that $K^{\mu}$ has dimensions of a length squared, and since it is describing a property of the secondary (its scalar dipole vector) it scales as $\mu^2$, with $\mu$ the characteristic lengthscale of the secondary. Hence $K^{\mu}(g,u,\Omega)\nabla_{\mu}\varphi \sim \mu \epsilon^2$, which is why we retain it. The direct implication is that all other terms like $\varphi \nabla \varphi, (\nabla\varphi)^2, \nabla\nabla\varphi,...$ are $O(\epsilon^3)$ or higher, and thus negligible. It then follows that a  scalar quadrupole tensor $D^{\mu\nu}$ associated to a $\nabla_{\mu} \nabla_{\nu} \varphi$ term is negligible, together with all the higher order multipoles.

Up until now, the discussion has been agnostic about the parity of the scalar field.
Let us first consider a parity even scalar field. As the Lagrangian needs to be parity even, the charges are $Q_{1,2}(g,u,\Omega) = Q_{1,2}(a_i)$, with $Q_{1,2}(a_i)$ generic functions of the $a_i$. For the dipole vector, the CH theorem implies that the most general form a vector built out of contractions of $u,\Omega$ can have is
\begin{equation}
\label{eq: generic vector out of u and Omega}
    K^{\mu}=\sum_{j=0}^{3} f_j(a_i)(\Omega^j u)^{\mu},
\end{equation}
with $f_j(a_i)$ generic functions of the $a_i$ that preserve an RPI Lagrangian. 

In the pseudoscalar case, let us write down the Lagrangian in terms of the pseudoscalar $\vartheta$ as
\begin{equation}
\label{eq: effective lagrangian with theta - theta^2 - grad theta}
    L_{\vartheta} = (\mathcal{Q}_1(g,u,\Omega) + \mathcal{Q}_2(g,u,\Omega)\vartheta)\vartheta + \mathcal{K}^{\mu}(g,u,\Omega)\nabla_{\mu}\vartheta.
\end{equation}
For the Lagrangian to be a parity even scalar, we require $\mathcal{Q}_1$ to be a pseudoscalar, $\mathcal{Q}_2$ to be a parity-even scalar as its multiplying $\vartheta^2$, and $\mathcal{K}^{\mu}$ to be a pseudovector.
Hence, the structure of the charges and the dipole vector in the parity-even and parity-odd cases are completely different.
As all the $a_i$ are parity-even scalars, $\mathcal{Q}_1(g,u,\Omega)$ cannot simply be a function of them. On the other hand, we still have $\mathcal{Q}_2(g,u,\Omega) = \mathcal{Q}_2(a_i)$ as this is the most general parity even scalar we can write down, as discussed. 

To construct $\mathcal{Q}_1$, we note that the most general pseudoscalar one can write down contracting $u,\Omega$, is
\begin{equation}
    \mathcal{Q}_1(g,u,\Omega) = (\star \Omega)^{\mu\nu}\Omega_{\mu\nu} f_4(a_i),
\end{equation}
where $(\star \Omega)^{\mu\nu} = \frac{1}{2}\epsilon^{\mu\nu\rho\sigma}\Omega_{\rho\sigma}$ is the dual of the angular velocity.
The underlying idea is that only one pseudoscalar can be written down: $(\star \Omega)^{\mu\nu}\Omega_{\mu\nu}$. This uniqueness follows from the identity $(\star\Omega)^{\alpha \sigma}\Omega_{\sigma \beta}  \sim (\star \Omega)^{\mu\nu}\Omega_{\mu\nu}\delta^{\alpha}_{\beta}$.
For the pseudovector, CH theorem yields
\begin{equation}
\label{eq: generic pseudovector out of u and Omega}
    \mathcal{K}^{\mu}=\sum_{j=0}^{3} g_j(a_i)(\Omega^j (\star\Omega u))^{\mu},
\end{equation}
with $g_j(a_i)$ preserving RPI. This expression highlights the presence of the fundamental pseudovector $\star \Omega^{\mu}_{\nu}u^{\nu}$, which is the kinematic counterpart of the spin vector $S^{\mu}=(\star S)^{\mu}_{\nu}u^{\nu}$. The presence of this pseudovector was to be expected, as it is well known that, in a dCS BH in isolation, its scalar dipole vector is proportional to the BH spin \cite{Yunes:2009hc}.

We note how this symmetry based construction shows how the spacelike dipole vectors, i.e. the vectors $K^\mu_{\perp}$, $\mathcal{K}^{\mu}_{\perp}$ are zero if there's no angular velocity. To have a dipole vector, one needs a privileged axis, provided in this case by rotation.

\section{Application to an EMRI with a secondary dressed by a scalar field}\label{sec: EMRI case}

We now apply the framework developed in the 
previous section to the case of EMRIs endowed 
with a scalar field. We build on the 
formalism developed in \cite{Maselli:2020zgv,Spiers:2023cva} 
for point particles, and follow Refs.~\cite{Spiers:2023cva,Mathews:2021rod} 
to consistently include self-force 
contributions.
We exploit the hierarchy of the two characteristic length-scales naturally embodied within EMRIs, 
given by the mass of the primary $M$, and of 
the secondary $\mu$, such that the mass ratio 
is $\epsilon=\mu/M \ll 1$. 
We also focus on theories described by 
the action \eqref{eq: general action} in 
which BH solutions evade no-hair theorems, and 
can develop nontrivial scalar field profiles.
In these models, deviations from GR affecting the 
primary and secondary BHs depend on the 
dimensionless parameter
\begin{equation}
    \zeta = \frac{\alpha}{M^n}, \quad  \zeta_p=\frac{\alpha}{\mu^n} \quad \rightarrow \quad \zeta = \epsilon^n \zeta_p\ .
\end{equation}
Given that the secondary in an EMRI is assumed to be 
a stellar mass compact object,  we conservatively assume  $\zeta_p \lesssim O(1)$, consistently with current GW and astrophysical bounds \cite{Nair:2019iur,Perkins:2021mhb,Sanger:2024axs,silva2021astrophysical,julie2025inspiral}. 
Then, the adimensional bGR parameter scales as
\begin{equation}
\label{eq:bGR-parameter-scaling}
    \zeta \sim O(\epsilon^n), \quad n \geq 2\ .
\end{equation}
As corrections to the primary first appear at $O(\zeta^2)$ (see e.g. \cite{Maselli:2020zgv}), it can be accurately modeled by the Kerr metric up to 
$O(\epsilon^4)$. The mass-ratio can be used as a single 
perturbative parameter to study the EMRI 
dynamics in this background.
Expanding the scalar in terms of $\epsilon$, 
we have
\begin{equation}
    \varphi= \varphi^{(0)} + \epsilon \varphi^{(1)}+O(\epsilon^2)=\varphi^{(0)}+\epsilon \phi,\label{scalar field expansion}
\end{equation}
$\varphi^{(0)}$ is the background scalar field, while $\phi\sim O(1)$. As the EFT is continuously connected to the GR limit as $\epsilon \rightarrow 0$, the no-hair theorem implies $\varphi^{(0)} = \text{const} + \zeta \phi_0$, where a possible correction to the trivial background appears at $O(\zeta) \equiv O(\epsilon^2)$, where the no-hair theorem can be circumvented. Without loss of generality 
we can fix the constant to zero, so that $\varphi^{(0)} = \epsilon^2 \phi_0$.

Similarly, we expand the metric tensor 
in the mass ratio:
\begin{align}
    \label{eq:metric expansion}
    g_{\mu\nu}=g^{(0)}_{\mu\nu}+\epsilon h^{(1)}_{\mu\nu}+\epsilon^2 h^{(2)}_{\mu\nu}+O(\epsilon^3) \ ,
\end{align}
where $g^{(0)}_{\mu\nu}$ corresponds 
to the Kerr metric. The perturbations 
$h_{\mu\nu}^{(n)}$ are sourced by the motion 
of the secondary, and push it away from 
its geodesic motion due to self-force 
effects \cite{Barack:2018yvs}. 
In what follows, we first neglect the direct 
contribution of $h^{(n)}_{\mu\nu}$, 
and reintroduce them in Sec.~\ref{sec: EMRI case - including self-force terms}.

\subsection{Mass-ratio expansion of the modified MPD equations}
\label{sec: EMRI case - expanding generalised MPD eqs}

In analogy with the small mass-ratio expansion of the 
field's equation, we can expand the modified MPD equations 
in terms of $\epsilon$.

Before doing so, it is instructive to analyze the 
dimensional scaling of Eqns.~\eqref{eq: generalised MPD equation - linear momentum}-\eqref{eq: generalised MPD equation - spin tensor}. 
As $\mathcal{P}^{\mu},\mathcal{S}_{\mu\nu},Q,K^{\mu}$ describe intrinsic properties of the secondary, 
we have
\begin{align}
    & \mathcal{P}/M \sim Q/M \sim O(\epsilon)\ , \\
   & \mathcal{S}/M^2 \sim K/M^2 \sim O(\epsilon^2)\ .
\end{align}
We then introduce the following notation: 
\begin{alignat}{3}
  \mathcal{S}_{\mu \nu} &= \mu^2  \hat{\mathcal{S}}_{\mu\nu}, &\qquad R^{\mu\nu\alpha\beta} &=\frac{\hat{R}^{\mu\nu\alpha\beta}}{M^2}, &\qquad L &=\mu \hat{L},
  \nonumber  \\ 
  \Omega_{\mu\nu} &= \frac{\hat{\Omega}_{\mu\nu}}{\mu}, &\qquad \mathcal{P}^{\mu} &= \mu \hat{\mathcal{P}}^{\mu}, &\qquad \nabla_{\mu} &= \frac{\hat \nabla_{\mu}}{M},
  \nonumber \\ 
  Q &= \mu \hat Q, &\qquad K^{\rho} &= \mu^2 \hat K^{\rho}, &\qquad \tau &= M \hat \tau,
  \nonumber \\
  G_{\mu\nu} &=  \frac{\hat G_{\mu\nu}}{M^2}, &\qquad \frac{\delta^{(4)}}{\sqrt{-g}} &= M^{-4}\frac{\hat \delta^{(4)}}{\sqrt{-\hat g}}.
  \label{eq:hat rules}
\end{alignat}
where, unless otherwise specified, 
$\hat{(\cdot)}$ quantities are 
dimensionless, and order $O(1)$.

Here, as $L$ is the matter Lagrangian describing 
the secondary, it is associated with the 
length-scale of the secondary. Similarly, 
$\Omega$ scales as $\mu^{-1}$ as describes 
the angular velocity of the secondary.

With this notation in hand, and using 
$\varphi = \epsilon \phi$, we proceed to 
expand the modified MPD equations in powers 
of the mass-ratio, neglecting terms 
$O(\epsilon^3)$ and higher. From 
Eqns.~\eqref{eq: generalised MPD equation - linear momentum}-\eqref{eq: generalised MPD equation - spin tensor} we have:
\begin{align}
    \label{eq: adimensional generalised MPD equation - linear momentum}
\frac{\mathrm{D}\hat{\mathcal{P}}^{\mu}}{\mathrm{d}\hat\tau} &= -\frac{\epsilon}{2} \hat R^{\mu \alpha \beta \gamma} \hat{\mathcal{S}}_{\beta \gamma}u_\alpha + \epsilon \hat{Q}_1(a_i)\hat\nabla^{\mu}\phi 
\nonumber \\ &
+ \epsilon^2 \hat{Q}_2(a_i)\phi \hat\nabla^{\mu}\phi
+ \epsilon^2 \hat{K}^{\rho}(a_i)\hat\nabla_{\rho}\hat\nabla^{\mu}\phi \ ,
\\ 
    \label{eq: adimensional generalised MPD equation - spin tensor}
   \epsilon \frac{\mathrm{D}\hat{\mathcal{S}}^{\mu\nu}}{\mathrm{d}\hat\tau} &= 2 \hat{\mathcal{P}}^{[\mu}u^{\nu]}+2 \epsilon^2 \hat K^{[\mu}\hat\nabla^{\nu]}\phi\ ,
\end{align}
where $\hat Q_1, \hat Q_2$ are defined from the expansion
\begin{equation}
\label{eq: scalar expansion of charge}
    \hat Q(a_i,b_j) = \hat Q_1(a_i) + \epsilon \phi \hat Q_2(a_i) + O(\epsilon^2)\ .
\end{equation}
Here, $\nabla_{\mu}\varphi$ and hence $b_2$ do not appear, as the term featuring it can be thought of as a monopole-dipole interaction, thus appearing at higher order in the mass-ratio. More precisely, we neglect the term $\epsilon \frac{\partial \hat Q}{\partial \nabla_{\mu}\varphi}\nabla_{\mu}\phi$, where the vector is a charge-dipole coupling with dimensions of a mass. As the vector is defined at the secondary, it scales with $\mu$, producing an extra $\epsilon$ term. Note also that the effect of the background scalar field $\epsilon^2 \phi_0$ is negligible, as it would produce an induced charge $\sim \epsilon^2 \phi_0 =O(\epsilon^2)$. 

Note that in Eq.~\eqref{eq: adimensional generalised MPD equation - linear momentum} we write 
$\hat{K}^{\mu}(a_i)$ instead of $\hat{K}(a_i,b_j)$. 
This approximation is justified as $K$ is 
order $\mathcal{O}(\epsilon^2)$, while 
$\varphi$ is $O(\epsilon)$. So any 
dependence of $K$ on $\varphi$ and 
$\nabla \varphi$ is higher order in 
the mass ratio and can be neglected. 
Exploiting this effective independence of 
$K^{\mu}$, and as $K^{\mu} = \frac{\partial L}{\partial b_2}\frac{\partial b_2}{\partial\nabla_{\mu}\varphi}$, we can write $b_2 = k^{\mu}(u,\Omega)\nabla_{\mu}\varphi$.

We now specialize to the case of a parity 
even scalar $\varphi$ which transforms as 
$\varphi \rightarrow \varphi$ under a parity 
inversion.
One can then apply the CH theorem to $k^{\mu}(u,\Omega)$ itself. Forbidding parity-odd scalars, the CH theorem implies that the most general form of $k^{\mu}(u,\Omega)$ is:
\begin{equation}
\label{eq: general dipole vector - little k}
    k^{\mu}=A(a_i)u^{\mu} + B(a_i)(\Omega u)^{\mu} + C(a_i)(\Omega^2 u)^{\mu}+ D(a_i)(\Omega^3 u)^{\mu},
\end{equation}
where $A,B,C,D$ are generic functions of the four scalars $a_i$, exactly as in Eq.~\eqref{eq: generic vector out of u and Omega} of \cref{subsec:scalar-vs-pseudoscalar}.

At this stage, the explicit form of $\hat Q(a_i,b_j), \hat K^{\mu}(a_i)$ is unknown. Our aim is to show that, at the mass-ratio order we are interested in, these quantities can be fully characterized by $u,\phi$ and unknown constant coefficients. 

Having assessed in the previous section the conservation of $\mathcal{S}_{\mu\nu}\mathcal{S}^{\mu\nu}$ along the worldline \eqref{eq: conservation of generalised S^2}, we want to check in this setting if we can say something about $\mathcal{P}^2$. We first rewrite 
Eq.~\eqref{eq: modified linear momentum - post generalised SSC} in the mass-ratio expansion, using dimensionless quantities
\begin{align}
\label{eq: adimensionalised modified linear momentum - post generalised SSC}
    (\hat{\mathcal{P}}u)\hat{\mathcal{P}}^{\mu} =& (\hat{\mathcal{P}}^2)u^{\mu} - \epsilon \hat{\mathcal{S}}^{\mu\nu}\frac{\mathrm{D}\hat{\mathcal{P}}_{\nu}}{\mathrm{d}\hat \tau}
    \nonumber \\ &
    +\epsilon^2\left[ (\hat{\mathcal{P}}\hat K)\hat{\nabla}^{\mu}\phi - (\hat{\mathcal{P}} \hat\nabla \phi )\hat{K}^{\mu} \right]\ .
\end{align}
Hence, given $\mathrm{D}\hat{\mathcal{P}}_{\mu}/\mathrm{d}\hat \tau = O(\epsilon)$, it follows that 
\begin{equation}
\label{eq: linear momentum proportional to four velocity at first order}
    \hat{\mathcal{P}}^{\mu} = \frac{\hat{\mathcal{P}}^2}{\hat{\mathcal{P}}u}u^{\mu} + O(\epsilon^2)\ ,
\end{equation}
and the generalized SSC \eqref{eq: generalised SSC} implies
\begin{equation}
\label{eq: spin four-velocity contraction from SSC}
    \hat{\mathcal{S}}^{\mu \nu}u_{\nu} = O(\epsilon^2)\ .
\end{equation}
Contracting Eq.~\eqref{eq: linear momentum proportional to four velocity at first order} with $u^{\mu}$, we also find
\begin{equation}
\label{eq: norm of generalised linear momentum}
    \hat{\mathcal{P}}^2 = -(\hat{\mathcal{P}}^{\mu}u_{\mu})^2 + O(\epsilon^2)\,
\end{equation}
which implies a timelike $\mathcal{P}^{\mu}$ as expected.
Now, consider the contraction of \eqref{eq: adimensionalised modified linear momentum - post generalised SSC} with $\mathrm{D}\hat{\mathcal{P}}_{\mu}/\mathrm{d}\hat \tau$. We get
\begin{equation}
    \frac{\mathrm{D}\hat{\mathcal{P}}^2}{\mathrm{d}\tau} = \frac{2\hat{\mathcal{P}}^2}{\hat{\mathcal{P}}u}u^{\mu}\frac{\mathrm{D}\hat{\mathcal{P}}_{\mu}}{\mathrm{d}\tau} + O(\epsilon)=O(\epsilon),
\end{equation}
where the last equality follows from the equation of motion for $\hat{\mathcal{P}}_{\mu}$ \eqref{eq: generalised MPD equation - linear momentum}. Hence, on the worldline
\begin{equation}
    \hat{\mathcal{P}}^2 = -\beta + O(\epsilon)\ ,
\end{equation}
with $\beta >0$ a constant. 
The former, together with 
Eqns.\eqref{eq: linear momentum proportional to four velocity at first order} and \eqref{eq: norm of generalised linear momentum} imply:
\begin{equation}
\label{eq: linear momentum equal to four velocity at first order}
    \hat{\mathcal{P}}^{\mu} = (\sqrt{\beta}  + O(\epsilon))u^{\mu} + O(\epsilon^2)= [1+O(\epsilon)]u^{\mu} + O(\epsilon^2),
\end{equation}
where in the last step we take the GR limit to fix the constant $\sqrt\beta = 1$, since 
our theory is continuously connected to 
GR. 

It is convenient to choose a tetrad vector, say $e^{\mu}_{I=0}$, aligned to the modified four-momentum $\mathcal{P}^{\mu}$, such 
that
\begin{equation}
\label{eq: gauge choice on tetrad bGR}
    e^{\mu}_{0}=\frac{\hat{\mathcal{P}}^{\mu}}{\sqrt{-\hat{\mathcal{P}}^2}}.
\end{equation}
Note that, in the GR limit, 
the convenient choice would be $e^{\mu}_{0} = p^{\mu}/\sqrt{-p^2}$, as in \cite{PhysRevD.73.104031}.
Expressing the angular velocity 
\eqref{eq: angular velocity and tetrad}
in terms of dimensionless quantities 
we have
\begin{equation}
    \hat\Omega^{\mu \nu}e_{\mu I} = \epsilon\frac{\mathrm{D}e^{\nu}_{I}}{\mathrm{d}\hat\tau}\ .
\end{equation}
and from the $I=0$ component and 
Eq.~\eqref{eq: gauge choice on tetrad bGR} 
we obtain the identity
\begin{equation}
    \hat\Omega^{\mu \nu}\hat{\mathcal{P}}_{\nu} = - \epsilon \frac{\mathrm{D}\hat{\mathcal{P}}^{\mu}}{\mathrm{d}\hat\tau}=O(\epsilon^2).
\end{equation}
Using Eq.~\eqref{eq: linear momentum equal to four velocity at first order}, it then follows that
\begin{equation}
\label{eq: auxiliary vector order mass-ratio and a3 order mass-ratio squared!}
    \hat\Omega^{\mu \nu}u_{\nu} = O(\epsilon^2) \quad \rightarrow \quad \hat a_3 = O(\epsilon^4).
\end{equation}
Physically, this means that the 'electric' or 'boost' part of $\Omega$ scales linearly with the mass-ratio.  

The implication of the above scaling is twofold. 
First, as $\hat a_3 \sim O(\epsilon^4)$, and hence $a_3 \sim O(\epsilon^2)$, 
all the generic functions of the scalars $a_i$ appearing 
in the equations of motion 
\eqref{eq: generalised MPD equation - linear momentum}-\eqref{eq: generalised MPD equation - spin tensor},  effectively 
depend on $a_2,a_4$ to the mass-ratio order 
we consider; i.e., neglecting terms of order $\mathcal{O}(\epsilon^3)$. Second, the dipole vector \eqref{eq: general dipole vector} and \eqref{eq: general dipole vector - little k} reads
\begin{equation}
    \label{eq: dipole vector proportional to four velocity}
    \hat{K}^{\mu} = \hat{K}(a_2,a_4)u^{\mu} + O(\epsilon),
\end{equation}
where $K(a_2,a_4)$ is the leading order term of $A(a_i) \partial L / \partial b_2$. As $\hat K^{\mu}$ always appears at second order in the mass ratio, for current purposes, the scalar dipole vector is simply proportional to the four-velocity. We will show that a dipole of this form does not affect the equations of motion.

Having shown that all the arbitrary functions 
in play only depend on $a_2,a_4$, we proceed 
to show how $a_2, a_4$ depend on $\phi$.
First, we note that it is natural to expect 
that quantities such as $a_2,a_4$ have an 
implicit dependence on the scalar field. 
Indeed, quantities such as the four-velocity and angular velocity must depend on $\varphi$, since the secondary's trajectory is influenced by the scalar field. 

Let us  decompose the angular velocity 
$\Omega_{\mu\nu}$ into its electric and 
magnetic components, such that 
\begin{equation}
    \Omega_{\mu \nu} = 2 u_{[\mu}E_{\nu]}+ \varepsilon_{\mu\nu\rho\sigma}u^{\rho}B^{\sigma}\ ,
\end{equation}
where 
\begin{equation}
\label{eq: electric-magnetic decomposition of the angular velocity}
    E_{\mu} = \Omega_{\mu\nu}u^{\nu}\quad \ , \quad B_{\mu} = -\frac{1}{2}\varepsilon_{\mu\beta\gamma\delta}\Omega^{\gamma \delta}u^{\beta}\ .
\end{equation}

Then, computing $a_2 = \Omega_{\mu\nu}\Omega^{\mu\nu}$, 
and using the condition \eqref{eq: auxiliary vector order mass-ratio and a3 order mass-ratio squared!}, 
we have:
\begin{equation}
    a_2 = 2 B_{\mu}B^{\mu} + O(\epsilon^2)\ .
\end{equation} 
Similarly, for $a_4$ 
\begin{equation}
\label{eq: a4 proportional to the square of a2}
    a_4 = 2 B^{\mu} B_{\mu} B^{\nu} B_{\nu} + O(\epsilon^2)= \frac{1}{2}a_2^2 + O(\epsilon^2).
\end{equation}
We have already shown that, along the worldline, $\mathcal{S}^{\mu\nu}\mathcal{S}_{\mu\nu}=c_1$ (see Eq.~\eqref{eq: conservation of generalised S^2}). 
Using the definition of $\mathcal{S}_{\mu\nu}$ 
in Eq.~\eqref{eq: generalized spin tensor - main tex}
yields
\begin{equation}
\label{eq: auxiliary equation with spin squared and F function}
    \hat{\mathcal{S}}^{\mu\nu}\hat{\mathcal{S}}_{\mu\nu} = F(a_2,\varphi) + O(\epsilon^2)=\hat{c}_1\ ,
\end{equation}
where $F$ is a function that depends on $a_2,\varphi$ via the Lagrangian and its derivatives with respect to $a_2,a_3,a_4$. The dependence on $a_{3,4}$ has been included in $O(\epsilon^2)$. 
We use $\varphi=\epsilon \phi$ and assume that $c_2$ is a constant satisfying $F(c_2,0)=\hat{c}_1$ such that $\partial_{a_2}F(c_2,0)\neq 0$. Assuming that $F$ is sufficiently regular near $(c_2,0)$, a first-order expansion of Eq.~\eqref{eq: auxiliary equation with spin squared and F function} on the branch for which $a_2 \rightarrow c_2$ as $\epsilon \rightarrow 0$, yields
\begin{equation}
\label{eq: a2 known up to epsilon squared terms}
    a_2=c_2+\epsilon\gamma_2\phi+O(\epsilon^2)\ ,
\end{equation}
with $\gamma_2$ the constant
\begin{equation*}
    \gamma_2 = -\frac{\partial_{\varphi}F(c_2,0)}{\partial_{a_2}F(c_2,0)}\ .
\end{equation*}

Using Eq.~\eqref{eq: a4 proportional to the square of a2} 
we also have
\begin{equation}
\label{eq: a4 known up to epsilon squared terms}
    a_4=c_4+\epsilon\gamma_4\phi+O(\epsilon^2)\ .
\end{equation}

Going back to the equation for the 
modified linear momentum, Eq.~\eqref{eq: adimensional generalised MPD equation - linear momentum}, we can finally characterize 
the unknown dependencies of $\hat Q_1(a_i), \hat Q_2(a_i), \hat K^{\rho}(a_i)$ on the 
$a_i$. 

Starting with the dipole vector, we use Eq.~\eqref{eq: dipole vector proportional to four velocity} and 
Eqns.~\eqref{eq: a2 known up to epsilon squared terms}-\eqref{eq: a4 known up to epsilon squared terms} to get
\begin{equation}
    \label{eq: dipole vector equal to four velocity times constant}
    \hat K^{\mu} = \hat K u^{\mu} + O(\epsilon)\ ,
\end{equation}
with $ \hat K$ a constant. Hence, 
\begin{equation}
\label{eq: physical dipole is negligible}
    \hat K^{\mu}_{\perp} = O(\epsilon).
\end{equation}
Following the discussion of 
Sec.~\ref{sec: effective action - multipoles and spin} 
on the ambiguity in defining $Q$ and 
$K^{\mu}$, we see that the imprint of the 
spacelike dipole vector $\hat K^{\mu}_{\perp}$ 
on the equations of motion is subdominant at the mass-ratio order we 
consider.

Similarly, for $\hat Q_1(a_i)$, using 
Eqns.~\eqref{eq: auxiliary vector order mass-ratio and a3 order mass-ratio squared!} and 
\eqref{eq: a2 known up to epsilon squared terms}-\eqref{eq: a4 known up to epsilon squared terms} 
we have
\begin{equation}
    \label{eq: characterization of Q1}
    \hat Q_1(a_i)=\hat Q_0 + \epsilon \phi \hat Q_{\chi} + O(\epsilon^2),
\end{equation}
where we have defined
\begin{align}
    \hat Q_0 &=\hat Q_1(c_2,c_4) ,
    \\ 
    \hat Q_{\chi} &= \left. \gamma_2 \frac{\partial \hat Q_1}{\partial a_2} \right|_{c_2,c_4} + \left. \gamma_4 \frac{\partial \hat Q_1}{\partial a_4} \right|_{c_2,c_4}.
\end{align}
and $\hat Q_0$ is the leading-order charge, 
and $\hat Q_{\chi}$ its correction 
induced by the spin-scalar field 
coupling. 
In a shift-symmetric Horndeski setting with a sGB term, $\hat Q_0$ would be related to the properties of the secondary $\mu,\chi$ via an integral over the horizon \cite{Saravani:2019xwx}.
The correction $\hat Q_{\chi}$ to the leading charge arises due to two effects:
on the one hand, the scalar field effectively changes the inertia of the body, similarly to how it defines an effective mass in the non-spinning case \eqref{eq: effective action - monopole}, hence affecting how the body spins. On the other hand, if the spin changes, the horizon is affected as well, and hence the charge.

Finally, for $\hat Q_2(a_i)$ we have
\begin{equation}
    \label{eq: characterization of Q2}
    \hat Q_2(a_i)=\hat Q_\varphi + O(\epsilon)\ ,
\end{equation}
where $\hat Q_{\varphi}$ is a constant which 
we interpret as a self-coupling-induced correction to the charge.

The equation for the modified linear momentum \eqref{eq: adimensional generalised MPD equation - linear momentum} can then be rewritten as 
follows
\begin{align}
    \label{eq: adimensional generalised MPD equation - linear momentum - almost final form}
    \frac{\mathrm{D}\hat{\mathcal{P}}^{\mu}}{\mathrm{d}\hat \tau} & = 
    - \frac{1}{2}\epsilon \hat R^{\mu \alpha \beta \gamma}\hat{\mathcal{S}}_{\beta \gamma}u_{\alpha} + \epsilon \hat Q_0 \hat \nabla^{\mu} \phi 
    \nonumber \\ &
  + \epsilon^2 \left(\hat Q_{\varphi \chi}\phi + \hat K u^{\rho}\hat\nabla_{\rho}  \right)\hat\nabla^{\mu}\phi\ ,
\end{align}
where we have defined $\hat Q_{\varphi \chi} = \hat Q_{\varphi} + \hat Q_{\chi}$. The equation of motion for the modified spin tensor is
\begin{equation}
\label{eq: adimensional generalised MPD equation - spin tensor - almost final form}
    \frac{\mathrm{D}\hat{\mathcal{S}}^{\mu \nu}}{\mathrm{d}\hat \tau} = 2 \epsilon^{-1} \hat{\mathcal{P}}^{[\mu}u^{\nu]} + 2 \epsilon \hat K u^{[\mu} \hat\nabla^{\nu]}\phi\ ,
\end{equation}
where $2 \epsilon^{-1} \hat{\mathcal{P}}^{[\mu}u^{\nu]} = O(\epsilon)$ due to \eqref{eq: adimensionalised modified linear momentum - post generalised SSC} (in agreement with the GR limit). We stress that $O(\epsilon^2)$ terms in the spin tensor equation are not needed as $\hat{\mathcal{S}}_{\mu\nu}$ appears in Eq.~\eqref{eq: adimensional generalised MPD equation - linear momentum - almost final form} multiplied by $\epsilon$.

The appearance of $\hat K$-dependent terms may be puzzling at first. We showed in \ref{sec: effective action - multipoles and spin} the existence of a redundancy in defining the charge and the dipole vector.
We fixed the redundancy by requiring a spacelike dipole vector. 
Similarly, we now show that the equations of motion are independent of the non-spacelike component $\hat K u^{\mu}$. 

Consider the definition of $\hat {\mathcal{P}}^{\mu}$ in terms of the effective Lagrangian, Eq.~\eqref{eq: generalized linear momentum - main tex}. Using Eq.~\eqref{eq: general dipole vector - little k} and \eqref{eq: dipole vector proportional to four velocity}, we have
\begin{equation}
    \label{eq: introducing physical momentum}
    \hat{\mathcal{P}}^{\mu} = \hat\Pi^{\mu} + \epsilon^2 \hat K \hat \nabla^{\mu} \phi,
\end{equation}
where $\Pi^{\mu} = \sum_i \frac{\partial L}{\partial a_i}\frac{\partial a_i}{\partial u^{\mu}}$. Plugging this decomposition 
into 
Eq.~\eqref{eq: adimensional generalised MPD equation - linear momentum - almost final form}-\eqref{eq: adimensional generalised MPD equation - spin tensor - almost final form}, every $\hat K$ dependence vanishes, and we get the equations
\begin{align}
    \label{eq: adimensional generalised MPD equation - linear momentum - final form}
    \frac{\mathrm{D}\hat{\Pi}^{\mu}}{\mathrm{d}\hat \tau}  = &
    - \frac{1}{2}\epsilon \hat R^{\mu \alpha \beta \gamma}\hat{\mathcal{S}}_{\beta \gamma}u_{\alpha} + \epsilon \hat Q_0 \hat \nabla^{\mu} \phi 
    \nonumber \\ &
  + \epsilon^2 \hat Q_{\varphi \chi}\phi\hat\nabla^{\mu}\phi\ ,
  \\ 
\label{eq: adimensional generalised MPD equation - spin tensor - final form}
    \frac{\mathrm{D}\hat{\mathcal{S}}^{\mu \nu}}{\mathrm{d}\hat \tau} =& 2 \epsilon^{-1} \hat{\Pi}^{[\mu}u^{\nu]}\ .
\end{align}
Once the SSC \eqref{eq: generalised SSC} is provided, alongside the field equations for $\phi$ and $g_{\mu\nu}$, the system of equations \eqref{eq: adimensional generalised MPD equation - linear momentum - final form}-\eqref{eq: adimensional generalised MPD equation - spin tensor - final form} forms a closed system that only depends on the constants $\hat{Q}_0, \hat{Q}_{\varphi\chi},\mu,M$ and the initial conditions.

\subsection{Modified MPD equations up to 1PA}\label{sec:modifiedMPD-1PA}

Up to now we have kept all terms up to $O(\epsilon^2)$ in all of our equations, including 2PA terms. We proceed to derive 1PA equations for the dynamics of the secondary by neglecting spin-spin terms. 
Such terms, of the form $S^{\mu\nu}S^{\alpha\beta}$, first enter at $O(\epsilon^2)$ as conservative corrections, and conservative $O(\epsilon^2)$ forces only affect the phase at 2PA \cite{Mathews:2021rod}. 

Consider now Eq.~\eqref{eq: adimensionalised modified linear momentum - post generalised SSC}. Using the equation of 
motion for the linear momentum \eqref{eq: adimensional generalised MPD equation - linear momentum - almost final form} 
at order $O(\epsilon)$, we have for $\hat{\mathcal{P}}^{\mu}$:
\begin{equation}
\label{eq: final form of modified linear momentum - neglecting spin-spin}
    \hat{\mathcal{P}}^{\mu} = -\hat m(\phi)u^{\mu} + \epsilon^2\left(\hat K(g^{\mu\nu}+u^{\mu}u^{\nu})+\hat Q_0 \hat{\mathcal{S}}^{\mu\nu}\right)\hat \nabla_{\nu}\phi\ ,
\end{equation}
where we have used, in agreement with \eqref{eq: generalized linear momentum - only scalar charge and no spin} in the nonspinning limit,
\begin{equation}
    \hat m (\phi) = -\frac{\hat{\mathcal{P}}^2}{\hat{\mathcal{P}}^{\mu}u_{\mu}}\ .
\end{equation}
Plugging Eq.~\eqref{eq: final form of modified linear momentum - neglecting spin-spin} into the equation of motion for 
the spin tensor \eqref{eq: adimensional generalised MPD equation - spin tensor - almost final form}, we obtain
\begin{equation}
    \label{eq: adimensional generalised MPD equation - spin tensor - final form - neglecting spin-spin}
    \frac{\mathrm{D}\hat{\mathcal{S}}^{\mu\nu}}{\mathrm{d}\hat\tau}=-2 \epsilon \hat Q_0 (\hat\nabla_{\rho}\phi) \hat{\mathcal{S}}^{\rho[\mu}u^{\nu]}.
\end{equation}
The source is a new term dependent on the scalar field that induces precession due to the linear coupling of the spin tensor with the gradient of the scalar field, and is proportional to the charge of the body.

Now we substitute Eq.~\eqref{eq: final form of modified linear momentum - neglecting spin-spin} back 
into the equation of motion for the linear 
momentum \eqref{eq: adimensional generalised MPD equation - linear momentum - almost final form} to obtain 
\begin{align}
\label{eq: auxiliary eqution of motion - with effective mass}
   & -\hat m(\phi) \frac{\mathrm{D}u^{\mu}}{\mathrm{d}\hat \tau} =   \frac{\mathrm{D}\hat m(\phi)}{\mathrm{d}\hat \tau}u^{\mu} - \frac{1}{2}\epsilon \hat{R}^{\mu \alpha \beta \gamma }\hat{\mathcal{S}}_{ \beta \gamma}u_{\alpha} + \epsilon \hat Q_0 \hat\nabla^{\mu}\phi 
    \nonumber \\ &
    + \epsilon^2 \left( \hat Q_{\varphi\chi}\phi \hat\nabla^{\mu}\phi - \hat Q_0 \hat{\mathcal{S}}^{\mu\nu}u^{\rho}\hat{\nabla}_{\rho}\hat{\nabla}_{\nu}\phi - \hat K u^{\mu} \frac{\mathrm{D}^2 \phi}{\mathrm{d}\hat\tau^2} \right)\ .
\end{align}
Contracting it with $u_{\mu}$ and neglecting $O(\epsilon^3)$ 
terms, we find 
\begin{equation}
\label{eq: expression for derivative of effective mass}
    \frac{\mathrm{D}\hat m(\phi)}{\mathrm{d}\hat \tau} = \epsilon \hat Q_0 \frac{\mathrm{D}\phi}{\mathrm{d}\hat \tau} + \epsilon^2 \hat{Q}_{\varphi \chi} \phi \frac{\mathrm{D} \phi}{\mathrm{d}\hat\tau} + \epsilon^2 \hat K \frac{\mathrm{D}^2 \phi}{\mathrm{d}\hat\tau^2}\ .
\end{equation}
Integrating in time, the former yields, 
neglecting $O(\epsilon^2)$ quantities:
\begin{equation}
\label{eq: expression for effective mass}
    \hat m (\phi) = -1 + \epsilon \hat Q_0 \phi\ ,
\end{equation}
where we fixed the integration constant to $-1$ 
to recover the GR limit, which is continuously 
connected to our theories.

Plugging Eq.~\eqref{eq: expression for derivative of effective mass} back into Eq.~ 
\eqref{eq: auxiliary eqution of motion - with effective mass} 
and using the relation ~\eqref{eq: expression for effective mass}, we obtain the equation of motion for the four velocity 
up to $O(\epsilon^2)$:
\begin{widetext}
\begin{align}
\label{eq: equation for the four velocity - final form - neglecting spin-spin terms}
    \frac{\mathrm{D}u^{\mu}}{\mathrm{d}\hat \tau} =- \frac{1}{2} \epsilon (1+\epsilon \hat Q_0 \phi) \hat R^{\mu \alpha \beta \gamma}\hat{\mathcal{S}}_{\beta \gamma}u_{\alpha}
    + \epsilon (\hat Q_0 + \epsilon(\hat Q_{\varphi \chi} + \hat Q^2_0)\phi ) (\hat \nabla^{\mu}\phi)_{\perp} 
    - \epsilon^2 \hat Q_0 \frac{\mathrm{D}}{\mathrm{d}\hat \tau}(\hat{\mathcal{S}}^{\mu\nu}\hat \nabla_{\nu}\phi).
\end{align}
\end{widetext}

The system of 1PA MPD equations \eqref{eq: adimensional generalised MPD equation - spin tensor - final form - neglecting spin-spin},\eqref{eq: equation for the four velocity - final form - neglecting spin-spin terms}, together with the field equations we derive next section, fully 
determines the motion of the spinning secondary, 
when spin-spin terms and self-force neglected. We stress that these equations are only valid if the extra field is a parity-even scalar. We will treat the pseudoscalar case explicitly in Sec.~\eqref{sec: EMRI case - pseudoscalar}.

Self-force effects can be readily implemented following \cite{Spiers:2023cva}, \cite{Mathews:2021rod}, and we explicitly add them in \ref{sec: EMRI case - including self-force terms}. The system depends only  on two unknown constants $\hat Q_0, \hat Q_{\varphi\chi}$, as well as the initial conditions.

It is easy to see that taking the GR limit ($\phi \rightarrow 0$) yields the same sets of equations as in \cite{Mathews:2021rod}, if self-force effects are neglected. Similarly, taking the nonspinning limit, we find that the source of \eqref{eq: equation for the four velocity - final form - neglecting spin-spin terms} reduces to the first- and second-order scalar-force introduced in \cite{Spiers:2023cva}, again modulo self-force effects (see Eqs.(38,46) therein). 

We can now revisit the scalar field correction to the point particle effective lagrangian. For a parity even scalar field, in Sec.~\ref{subsec:scalar-vs-pseudoscalar}, we derived the lagrangian \eqref{eq: effective lagrangian with phi - phi^2 - grad phi} using a gradient expansion motivated by the EMRI limit. It features two charges $Q_{1,2}(a_i)$ and scalar dipole vector $K^{\mu}(a_i)$, whose dependence on the $a_i$ was unknown. In this section, we have shown that to derive 1PA equations this dependence is irrelevant, and the two charges can be treated as constants. Moreover, the scalar dipole vector does not enter the equations at 1PA. Hence, we can rewrite the Lagrangian in Eq.~\eqref{eq: effective lagrangian with phi - phi^2 - grad phi} as 
\begin{equation}
\label{1PAlangrangian}
    L_{\varphi}^{1\text{PA}} = Q_0 \varphi + \frac12 Q_{\varphi \chi} \varphi^2 .
\end{equation}

\subsection{Mass-ratio expansion of the scalar and gravitational field equations}
\label{sec: EMRI case - expanding field equations}

We now turn to expanding the Einstein and 
scalar field equations in powers of 
the mass ratio, isolating contributions up 
to $\mathcal{O}(\epsilon^2)$ order. 
Consider first equation \eqref{eq: scalar field equations - generic} for $\varphi$. As discussed in \cite{Spiers:2023cva}, for a parity-even scalar field, the source due to the coupling action $\alpha S_c$ only first contributes at $O(\epsilon^2)$ as follows:
\begin{equation}
    \label{eq: scalar source from gauss bonnet invariant}
    \hat\alpha_{GB} \hat\Sigma_c = - \epsilon^2 \hat \alpha_{GB}\hat{\mathcal{G}}^{(0)} + O(\epsilon^3),
\end{equation}
where $\mathcal{G}^{(0)}$ is the Gauss-Bonnet invariant of the background Kerr spacetime, $\hat{\mathcal{G}} = M^4 \mathcal{G}$, and $\alpha_{GB}$ is a coupling constant: $[\alpha_{GB}] = M^{2}$, $\alpha_{GB}/M^2 = O(\epsilon^2)$ in agreement with \eqref{eq:bGR-parameter-scaling}. It follows that $\hat \alpha_{GB} = \alpha_{GB}/\mu^2 = O(1)$. This term comes from the nonminimal linear coupling $\alpha_{GB} \varphi \mathcal{G}$ appearing in the action $\alpha S_c[g,\varphi]$:
\begin{equation}
    \alpha S_c \supset \alpha_{GB} S_c^{GB} = \frac{\alpha_{GB}}{16 \pi} \int \mathrm{d}^{4}x \sqrt{-g} f(\varphi)\mathcal{G}. 
\end{equation}

The dimensionless pole-dipole 
distribution sourcing the scalar 
field equation reads
\begin{align}
    \hat \Sigma =& -16 \pi \epsilon \int \mathrm{d}\hat\tau \hat Q(a_i,b_j) \frac{\hat\delta^{(4)}(x-z(\tau))}{\sqrt{-\hat g}}
    \nonumber \\ &
    +16 \pi \epsilon^2 \hat\nabla_{\mu}\int \mathrm{d}\hat\tau \hat K^\mu  \frac{\hat\delta^{(4)}(x-z(\tau))}{\sqrt{-\hat g}}\ .
\end{align}
We can then use the leading-order contribution 
for $\hat K^{\mu}$ in Eq.~\eqref{eq: dipole vector equal to four velocity times constant}, given 
the dipole component orthogonal to the four-velocity 
would contribute with an $O(\epsilon^3)$ term.
Integrating by parts and neglecting boundary terms, 
we obtain
\begin{equation}
    \label{eq: scalar field equation - mass-ratio expanded}
   \begin{split}
   \hat\Box \phi =&  - \epsilon \hat \alpha_{GB}\hat{\mathcal{G}}^{(0)} \\
   &-16 \pi  \int \mathrm{d}\hat\tau \frac{\hat\delta^{(4)}(x-z(\tau))}{\sqrt{-\hat g}} \left(\hat Q_0 + \epsilon \phi \hat Q_{\varphi\chi} \right).
   \end{split}
\end{equation}
Let's discuss some properties of the scalar field 
equation. First, the dipole vector does not affect 
the dynamics of $\phi$. 
Furthermore, the 
contribution of the modified background given by 
the first term in the r.h.s. 
of Eq.~\eqref{eq: scalar field equation - mass-ratio expanded} is time-independent. 
Thus, the dynamics of a parity-even scalar field 
is completely captured by the charge $Q_0$ and by the constant $Q_{\varphi\chi}$, with the latter 
encoding the strength of the 
charge-spin coupling and 
charge-scalar-field self-coupling.

Moreover, we note that the 
scalar field $\phi$ appearing 
within the integral in 
Eq.~\eqref{eq: scalar field equation - mass-ratio expanded} is a solution 
of the lower order equation $\hat\Box \phi =  -16 \pi  \int \mathrm{d}\hat\tau \hat\delta^{(4)}\hat Q_0 $, 
and hence proportional to the monopolar charge $Q_0$. 
This also shows that the integral 
vanishes if the secondary has 
$\hat Q_0 = 0$, as is the case for a neutron star in shift-symmetric sGB gravity \cite{Yagi:2015oca}.

Let us now perform the mass-ratio expansion of the Einstein equations \eqref{eq: Einstein equation for general action}. As shown in Ref.~\cite{Spiers:2023cva}, $T_{\mu\nu}^c$ first appears at order 
$O(\epsilon^3)$, hence is negligible. 
From Eq.~\eqref{eq: stress-energy tensor with scalar dipole correction},
we also see that the term 
$K^{\mu}\nabla_{\mu} \varphi$ 
entering the secondary stress-energy tensor 
$T_{\mu\nu}^p$ provides a 
$O(\epsilon^3)$ correction, 
as $\nabla \varphi \sim O(\epsilon)$. 
The remaining terms contribute such that 
the gravitational field equations become
\begin{align}
    \label{eq: gravitational field equations - mass-ratio expanded}
    \hat G_{\mu\nu} =& 8 \pi \epsilon \int \mathrm{d}\hat\tau \frac{\hat\delta^{(4)}(x-z(\tau))}{\sqrt{-\hat g}} \hat{\mathcal{P}}_{(\mu}u_{\nu)}
    \nonumber \\ &
    -8 \pi \epsilon^2 \hat\nabla^{\rho}\int \mathrm{d}\hat\tau \frac{\hat\delta^{(4)}(x-z(\tau))}{\sqrt{-\hat g}} \hat{\mathcal{S}}_{\rho(\mu}u_{\nu)}
    \nonumber \\ &
    +\frac{1}{2}\epsilon^2\hat\nabla_{\mu}\phi \hat\nabla_{\nu}\phi - \frac{1}{4}\epsilon^2 g_{\mu\nu}\hat\nabla^{\rho} \phi \hat\nabla_{\rho} \phi \ .
\end{align}

We now proceed to neglect spin-spin terms to obtain the 1PA field equations. We can exploit 
Eq.~\eqref{eq: final form of modified linear momentum - neglecting spin-spin} and Eq.~\eqref{eq: expression for effective mass} 
to simplify the gravitational field equations \eqref{eq: gravitational field equations - mass-ratio expanded}:
\begin{widetext}
    \begin{align}
    \label{eq: gravitational field equations - mass-ratio expanded - neglecting spin-spin terms}
    \hat G_{\mu\nu} =& 8 \pi \epsilon \int \mathrm{d}\hat\tau \frac{\hat\delta^{(4)}(x-z(\tau))}{\sqrt{-\hat g}} (1 - \epsilon \hat Q_0 \phi)u_{\mu}u_{\nu}
    -8 \pi \epsilon^2 \hat\nabla^{\rho}\int \mathrm{d}\hat\tau \frac{\hat\delta^{(4)}(x-z(\tau))}{\sqrt{-\hat g}} \hat{\mathcal{S}}_{\rho(\mu}u_{\nu)}
    +\frac{1}{2}\epsilon^2\hat\nabla_{\mu}\phi \hat\nabla_{\nu}\phi - \frac{1}{4}\epsilon^2 g_{\mu\nu}\hat\nabla^{\rho} \phi \hat\nabla_{\rho} \phi \ .
\end{align}
\end{widetext}
Equations \eqref{eq: scalar field equation - mass-ratio expanded} and \eqref{eq: gravitational field equations - mass-ratio expanded - neglecting spin-spin terms} are the complete 1PA field equations.

\subsection{The perturbative self-force expansion}
\label{sec: EMRI case - including self-force terms}

In the previous sections, the singular and regular split of the metric perturbation and scalar perturbations has been implicit in our analysis. Without such a split, equations like \cref{eq: scalar charge distribution - nonspinning particle with charge}, which contain products of Dirac-delta functions with singular quantities on the worldline, become ill-defined. In this section, we define the singular and regular split and express the split explicitly in the field equations and equations of motion from \cref{sec: EMRI case}. We also apply a mass-ratio expansion to express the field equations and equations of motion at first and second order.

Detweiler--Whiting~\cite{detwhit2003} defined the singular--regular split in GR for both the metric tensor and a scalar field. The singular (${\cal S}$) and regular (${\cal R}$) split is expressed as: 
\begin{align}
    h^{(n)}_{\mu\nu}=h^{(n)\S}_{\mu\nu} +h^{(n)\R}_{\mu\nu}, \\
    \varphi^{(n)}=\varphi^{(n)\S} + \varphi^{(n)\R}.
\end{align}
The regular parts of the field perturbations $h^{\R}_{\mu\nu}$ and $\varphi^{\R}$ are smooth along the worldline, not sourced by the small body, and give rise to the self-force~\cite{detwhit2003,dirac1938classical}. The singular fields are local particular solutions sourced by the stress-energy on the worldline and fixed by the body's multipole moments, and they exert no force on the body. Ref.~\cite{detwhit2003} derives the singular and regular split of both a scalar field and a gravitational field to linear order. The singular and regular split has been generalized to the non-linear fields in GR~\cite{harte2012mechanics, pound2012field, harte2015motion}. Our perturbative approach on a GR background generalizes these results to scalar-tensor theories of gravity, as in Refs.~\cite{Spiers:2023cva, zimmerman2015gravitational}. 
At first order in the mass ratio, the metric and scalar perturbations obey decoupled equations, namely the linearised Einstein equation and a minimally coupled massless scalar wave equation on the GR background, so the Detweiler--Whiting split applies directly~\cite{detwhit2003}, and the first-order motion is governed by the regular fields~\cite{zimmerman2015gravitational}. Beyond first order, the scalar-tensor field equations contain nonlinear metric-scalar couplings not covered by Refs.~\cite{harte2012mechanics, pound2012field, harte2015motion}; there we assume that an analogous split exists, with the effective fields $(\tilde g_{\mu\nu},\tilde\varphi)$ regular on the worldline, as in Ref.~\cite{Spiers:2023cva}.
To express the mass-ratio expansion, we define
\begin{align}\label{eq:regular}
h^{\R}_{\mu\nu}=\sum_{n=1} \epsilon^n h^{\R(n)}_{\mu\nu}\quad,\quad
\varphi^{\R}=\sum_{n=1} \epsilon^n \varphi^{(n)}_\R.
\end{align}

Using our definition of the regular field in Eqns.~\eqref{eq:regular} we define an effective metric
 and an effective scalar field,
\begin{align}
\tilde{g}_{\mu\nu}=g^{(0)}_{\mu\nu}+h^{\R}_{\mu\nu}\quad, \quad
    \tilde{\varphi}=\varphi^{\R}\ ,\label{eq:tildephiexpansion}
\end{align} 
which holds as we absorb $\varphi^{(0)}$ into the perturbations (following the discussion below \cref{scalar field expansion}). First- and second-order self-force calculations in GR indicate that compact spinless objects follow motion equivalent to that of a test body in the effective metric~\cite{misata1997, quwa1997, pound2012second, gralla2012}, as generalized to scalar-tensor theories in Ref.~\cite{Spiers:2023cva}. For spinning compact bodies in GR, a spin-curvature force pushes the compact object off geodesic motion~\cite{Mathisson2010,Dixon:1974xoz,papapetrou1951spinning}.

We can apply, under the ansatz of Ref.~\cite{Spiers:2023cva}, the singular and regular split to the particle action, \cref{eq: effective action - spin and pole-dipole}, 
\begin{equation}
    \label{eq: effective action - dipole}
    S_p=\int \mathrm{d}\ttau L(\tilde a_i,\tvarphi,\tilde k^{\mu}\tilde\nabla_{\mu}\tvarphi).
\end{equation}
where $\tilde\nabla$ is the covariant derivative in the effective metric, $\tilde{u}^\alpha:=dz^\alpha/d\tilde{\tau}$, with $\tilde{\tau}$ being the proper time in the effective metric, the scalars $\tilde a_i$ are functions of $\tilde g,\tilde u,\tilde \Omega$, with $\tilde\Omega_{\mu \nu} := \tilde e^{I}_{\mu}\frac{\tilde D}{d \ttau}\tilde e_{\nu I}= \tilde e^{I}_{\mu}\tilde u^\gamma\tilde\nabla_\gamma \tilde e_{\nu I}$, and $\tilde k^\mu$ is also a function of $\tilde g,\tu,\tilde\Omega,\tvarphi,\tilde\nabla\tvarphi$. The remaining parts of the action~\eqref{eq: general action}, $S_0[g,\varphi]$ and $S_c[g,\varphi]$, remain functions of the full fields in the full spacetime metric.

The derivation in the preceding sections changes trivially, with the corresponding quantities being explicitly labeled as being in the effective metric with a tilde. For clarity and brevity, we provide only the field equations and equation of motion of the secondary with explicit effective-metric labeling. 

\subsubsection{Scalar self-force equations\label{sec:scalar-self-force-equations}}

As in Ref.~\cite{Spiers:2023cva}, we split the equation of motion into scalar and gravitational contributions and expand in powers of $\epsilon$. We diverge from Ref.~\cite{Spiers:2023cva} by including spin and spin-scalar coupling terms for the acceleration. We define the acceleration in the effective metric as
\begin{align}\label{eq:acceleration}
    \tilde a^\mu:= \frac{\tilde D\tilde u^{\mu}}{\mathrm{d}\tilde\tau} = \tilde u^\nu \tilde\nabla_\nu \tilde u^\mu = \tilde a^\mu_{\rm spin} + \tilde a^\mu_{\rm scal}+\tilde a^\mu_{\rm spin\times scal}.
\end{align}
where $\tilde a^\mu_{\rm spin}$, $\tilde a^\mu_{\rm scal}$, and $\tilde a^\mu_{(2)\rm \rm{spin\times scal}}$ are the acceleration terms from the spin, scalar self-force, and spin-scalar coupling, respectively. For practical applications, and to relate to the equations of motion derived in the previous sections, we write the scaled acceleration ($\hat a^\mu=M a^\mu$) in the background metric as
\begin{align}
    \hat a^\mu:&= \frac{\mathrm{D} u^{\mu}}{\mathrm{d}\hat \tau} =  u^\nu \hat \nabla_\nu  u^\mu \notag \\
    &= \epsilon\left(\hat a^\mu_{(1)\rm grav}+  \hat a^\mu_{(1)\rm spin} + \hat a^\mu_{(1)\rm scal}\right)   \notag \\
    &\quad + \epsilon^2\left(\hat a^\mu_{(2)\rm grav}+  \hat a^\mu_{(2)\rm spin}+ \hat a^\mu_{(2)\rm scal}+\hat a^\mu_{(2)\rm spin\times scal}\right) \notag \\
    &\quad +O(\epsilon^3) .\label{eq:acc-background}
\end{align}
Here we converted to the scaled proper time and the covariant derivative in the background spacetime, as defined in \cref{eq:hat rules}. 

Next we define the acceleration terms in \cref{eq:acc-background} by expanding \cref{eq: equation for the four velocity - final form - neglecting spin-spin terms}; we convert the relevant quantities to the effective metric and scalar field \cref{eq:tildephiexpansion} and implement the $\epsilon$ expansion in \eqref{scalar field expansion} and \eqref{eq:metric expansion}. Expanding the left-hand side of \cref{eq: equation for the four velocity - final form - neglecting spin-spin terms} using \cref{eq:tildephiexpansion,eq:regular}, we find the first-order gravitational self-force to be identical to GR,
\begin{equation}\label{eq:a1grav}
     \hat a^\mu_{(1)\rm grav} =  -\frac{1}{2} P^{\mu\nu}(2\hat \nabla_\gamma h^{(1)\R}_{\nu \delta}-\hat \nabla_\nu h^{(1)\R}_{\delta \gamma})u^\delta u^\gamma\ ,
\end{equation}
with $P^{\mu\nu}=g^{\mu\nu}_{(0)} +u^\mu u^\nu$. From the same expansion, the second-order gravitational self-force with respect to the background spacetime also takes the same form as GR,
\begin{align}\label{eq:a2grav}
     \hat a^\mu_{(2)\rm grav} &=  -\frac{1}{2} \Big[P^{\mu\nu}(2\hat \nabla_\gamma h^{(2)\R}_{\nu \delta}-\hat \nabla_\nu h^{(2)\R}_{\delta\gamma})  \notag \\ & \ \ \ \ \ - P^{\mu\nu}h_{\nu(1)\mathcal{R}}^{\  \tau}(2\hat \nabla_\gamma h^{(1)\R}_{\tau \delta}-\hat \nabla_\tau h^{(1)\R}_{\delta\gamma}) \Big] u^\delta u^\gamma , 
\end{align}
but we note that deviations from GR appear in $h^{(2)\R}_{\delta\gamma}$ (as we will find from \cref{eq:dGh2}). 

Expanding the spin terms in \cref{eq: equation for the four velocity - final form - neglecting spin-spin terms} (while setting $\hat Q_0=0$) using \cref{eq:tildephiexpansion,eq:regular} gives the first-order spin-force,
\begin{equation}\label{eq:a1spin}
    \hat a_{(1)\rm spin}^\mu =  -\frac{1}{2}   \hat R_{(0)}^{\mu \alpha \beta \gamma}\hat{\mathcal{S}}_{\beta \gamma}u_{\alpha}\ ,
\end{equation}
which is identical to the spin-force term in the self-force expansion of the MPD equations in GR.
Note that while \cref{eq:a1spin} is first order in $\epsilon$, it does not contribute to 0PA dynamics as it is conservative. The same expansion gives the second-order spin-induced force,
\begin{align}\label{eq:a2spin}
    \hat a_{(2)\rm spin}^\mu &=  +\frac{1}{4}h^{(1)\R}_{\rho\tau}u^\rho u^\tau \hat R_{(0)}^{\mu \alpha \beta \gamma}\hat{\mathcal{S}}_{\beta \gamma}u_{\alpha}\notag \\ 
    & +\frac{1}{2}P^{\mu\nu}(2\hat \nabla_\gamma \hat \nabla_{(\beta}
    h^\R_{\alpha)\nu}-\hat \nabla_\gamma \hat \nabla_{\nu}
    h^\R_{\alpha\beta}\Big)\hat S^{\beta\gamma}u^\alpha,
\end{align}
which takes the same form as in GR. The evolution of the spin tensor is given by the expansion of  \cref{eq: adimensional generalised MPD equation - spin tensor - final form - neglecting spin-spin} using \cref{eq:tildephiexpansion,eq:regular}, which gives
\begin{align}
    \label{eq: adimensional generalised MPD equation - spin tensor - final form - neglecting spin-spin - SF}
    \frac{\mathrm{D}\hat{\mathcal{S}}^{\mu\nu}}{\mathrm{d}\hat\tau}&=\epsilon\Big(u^{(\rho} \hat{\mathcal{S}}^{\sigma)[\mu}g_{(0)}^{\nu]\lambda}(2 \nabla_{\sigma}
    h^\R_{\lambda\rho}-\nabla_{\lambda}
    h^\R_{\rho\sigma}  ) \notag \\ 
    & \quad \quad-2  \hat Q_0 (\hat\nabla_{\rho}\varphi) \hat{\mathcal{S}}^{\rho[\mu}u^{\nu]} \Big)+O(\epsilon^2).
\end{align}
\cref{eq: adimensional generalised MPD equation - spin tensor - final form - neglecting spin-spin - SF} is equivalent to the GR form plus a scalar charge ($Q_0$) dependent term that enters at order $\epsilon$. We neglect $O(\epsilon^2)$ corrections to the spin as they are conservative and so contribute to 2PA order.

Expanding the scalar terms in \cref{eq: equation for the four velocity - final form - neglecting spin-spin terms} (while setting $\hat{\mathcal{S}}^{\mu\nu}=0$) using \cref{eq:tildephiexpansion,eq:regular} gives the first- and second-order scalar accelerations,
\begin{equation}\label{eq:a1scal}
    \hat a_{(1)\rm scal}^\mu =\hat Q_0 P^{\mu\nu}\hat\nabla_\nu \varphi^{(1)}_\R \ .
\end{equation}
and
\begin{align}
    \hat a_{(2)\rm scal}^\mu &=  P^{\mu\nu}\Big(\hat Q_0\hat \nabla_\nu  \varphi^{(2)}_{\R} +(\hat Q^2_0 + \hat Q_{\varphi \chi})\varphi^{(1)}_{\R}\nabla_\nu   \varphi^{(1)}_{\R} \Big) \notag \\ 
    & \ \ \ - \hat Q_0 P^\mu_{\ \lambda}\big(h^{\lambda\nu}_{\mathcal R(1)}+h^{\mathcal R(1)}_{\rho\sigma}u^\rho u^\sigma g^{\lambda\nu}_{(0)}\big)\nabla_\nu \varphi^{(1)}_{\mathcal{R}},\label{eq:a2scal}
\end{align}
respectively. Their form is similar to the purely monopole scalar equations in Ref.~\cite{Spiers:2023cva}, where $m_{[1]}\rightarrow-\mu\hat Q_{0}$, and $m_{[2]}\rightarrow-\mu\hat Q_{\varphi \chi}/2$.

The final acceleration term comes from a novel coupling between the spin tensor and the scalar field. Expanding the remaining $\hat Q_0 \hat{\mathcal{S}}^{\mu\nu}$ term in \cref{eq: equation for the four velocity - final form - neglecting spin-spin terms}  using \cref{eq:tildephiexpansion,eq:regular}, gives
\begin{align}
    \hat a^\mu_{(2)\rm spin\times scal}=&-\hat Q_0 \frac{D}{\mathrm{d}\hat \tau}(\hat{\mathcal{S}}^{\mu\nu}\hat \nabla_{\nu}\varphi^{(1)}_{\mathcal R})
    \nonumber \\ &
    -\frac12\hat Q_0\,\varphi^{(1)}_{\mathcal R}\,\hat R^{\mu\alpha\beta\gamma}_{(0)}\hat{\mathcal{S}}_{\beta\gamma}u_\alpha,
\end{align}
while $\hat a^\mu_{(1)\rm spin\times scal}=0$.

We next express the field equations for the metric and scalar fields perturbatively. To do so, we require expanding the nonlinearity of the Einstein tensor~\cite{pound2022black, Barack:2018yvs}: for a generic tensor $g^{(0)}_{\mu\nu}+x_{\mu\nu}$, the expansion can be expressed in orders of the input, 
\begin{align}\label{eq:Gexpansion}
G_{\mu\nu}[x_{\alpha\beta}]= \delta G_{\mu\nu}[x_{\alpha\beta}]&+\delta^2 G_{\mu\nu}[x_{\alpha\beta},x_{\alpha\beta}]\notag \\ &+\delta^3 G_{\mu\nu}[x_{\alpha\beta},x_{\alpha\beta},x_{\alpha\beta}]+\ldots
\end{align}
where 
\begin{equation}
\delta^n G = \frac{1}{n!}\frac{d^n}{d\lambda^n}G\left[g^{(0)}_{\alpha\beta}+\lambda x_{\alpha\beta}\right]\big|_{\lambda=0}\ .
\end{equation}
The Einstein field tensor can then be expanded as
\begin{align}\label{eq:expandedEFE}
G_{\mu\nu}[g_{\alpha\beta}]&=\epsilon \delta G_{\mu\nu}[h^{(1)}_{\alpha\beta}] \notag \\ & \ \ \ + \epsilon^2 \Big(\delta G_{\mu\nu}[h^{(2)}_{\alpha\beta}]+\delta^2G_{\mu\nu}[h^{(1)}_{\alpha\beta},h^{(1)}_{\alpha\beta}] \Big) +\mathcal{O}(\epsilon^3). \notag \\
\end{align}
Expanding \cref{eq: gravitational field equations - mass-ratio expanded - neglecting spin-spin terms}, the first-order equation for the metric perturbation is therefore
\begin{equation}\label{eq:dGh1}
    \delta \hat G_{\mu\nu}[h^{(1)}_{\alpha\beta}]= 8\pi \int_\gamma \frac{\hat\delta^{4}[x^\mu-z^\mu[\tau]]}{\sqrt{-g^{(0)}}}u_\mu u_\nu d\hat\tau\ ,
\end{equation}
and the second-order equation is,
\begin{align}
    & \delta \hat G_{\mu\nu}[h^{(2)}_{\alpha\beta}]=-\delta^2\hat G_{\mu\nu}[h^{(1)}_{\alpha\beta},h^{(1)}_{\alpha\beta}]+\frac{1}{2}\hat \nabla_\mu\varphi^{(1)} \hat \nabla_\nu \varphi^{(1)}\notag \\
    &-\frac{g^{(0)}_{\mu\nu}}{4}\hat \nabla_\gamma\varphi^{(1)}\hat \nabla^\gamma \varphi^{(1)} \notag \\
    &+ 4\pi \int_\gamma d\hat \tau \frac{\hat \delta^{4}[x^\alpha-z^\alpha[\tau]]}{\sqrt{-g^{(0)}}}\bigg(-2\hat Q_0\varphi^{(1)}_\R u_\mu u_\nu \notag \\
    &+  \big( 4h^{\R(1)}_{\tau(\mu}u_{\nu)}u^\tau -u_\mu u_\nu (g_{(0)}^{\tau\delta}-u^\tau u^\delta)h^{\R(1)}_{\tau\delta}\big) \bigg)  \ \notag \\
    & -8 \pi  \hat\nabla^{\rho}\int \mathrm{d}\hat\tau \frac{\hat\delta^{(4)}(x-z(\tau))}{\sqrt{-g^{(0)}}} \hat{\mathcal{S}}_{\rho(\mu}u_{\nu)}.\label{eq:dGh2}
\end{align}
\cref{eq:dGh2} contains scalar monopole contributions (as in Ref.~\cite{Spiers:2023cva}) and spin contributions equivalent to those in GR~\cite{Mathews:2021rod}.

The final set of equations in our self-force expansion are for the perturbations to the scalar field. Expanding \cref{eq: scalar field equation - mass-ratio expanded}, the first-order scalar field equation is
\begin{align}\label{eq:first-orderscalarwave}
    \hat \Box \varphi^{(1)} &= -16\pi \int_\gamma \hat Q_0\frac{\hat \delta^{(4)}[x^\mu-z^\mu[\tau]]}{\sqrt{-g}}d\hat \tau\ , 
\end{align}
as in Refs.~\cite{Damour:1992we, Maselli:2020zgv, Spiers:2023cva}. The expansion to second order gives
\begin{align}
    \hat \Box \varphi^{(2)} &=  -\hat\alpha_{GB}\,\hat{\mathcal G}^{(0)} + h^{\mu\nu}_{(1)} \hat \nabla_\mu \hat\nabla_\nu \varphi^{(1)} +(\hat\nabla^\mu  h^{(1)}_{\mu\nu})\hat\nabla^\nu \varphi^{(1)} \notag \\
    & \ \ \ -\frac{1}{2} (\hat\nabla^\nu  h_{(1)}) \hat\nabla_\nu  \varphi^{(1)}-16\pi  \int_\gamma \Big[ \hat Q_{\varphi\chi}\varphi^{(1)}_R  \notag \\
    & \ \ \  -\frac{1}{2}  \hat Q_0 P^{\mu\nu}h_{\mu\nu}^{R(1)} \Big] \frac{\hat\delta^{(4)}[x^\alpha-z^\alpha[\tau]]}{\sqrt{-g^{(0)}}} d\hat\tau  ; \label{eq:second-orderscalarwave}
\end{align}
which, similarly to \cref{eq:a2scal}, is of the form of the second-order scalar field equation for the pure scalar monopole in Ref.~\cite{Spiers:2023cva} with the addition of the spin degree of freedom, $\hat Q_{\chi}$, absorbed into $\hat Q_{\varphi\chi}$ due to a degeneracy. 

\section{Secondary dressed by a pseudoscalar}
\label{sec: EMRI case - pseudoscalar}

We now determine the modified MPD equations 
of motion for a spinning particle, and 
the corresponding field's equations, in the case of a pseudoscalar $\vartheta$.
We build on the EFT approach described 
in Sec.~\ref{sec: EMRI case}. The action now reads
\begin{equation}
\label{eq: general action - pseudoscalar}
    S[g,\vartheta] = S_0[g,\vartheta] + \alpha S_c[g,\vartheta]+S_p[g,\vartheta] \ .
\end{equation}
Considering again theories with $[\alpha]=[mass^n], n \geq 2$, we can carry 
over here the mass-ratio analysis 
performed in Sec.~\ref{sec: EMRI case} 
for the parity-even scalar, in order 
to evaluate the contribution of 
$\alpha S_c[g,\vartheta]$ to the 
field equations. 
Hence, the only non-negligible nonminimal coupling is characterized by a coupling constant $[\alpha_{\text{dCS}}]=M^2$, and reads (see e.g. \cite{Yunes:2009hc})
\begin{equation}
    \alpha S_c \supset \alpha_{\text{dCS}}S_c^{\text{dCS}}=\frac{\alpha_{\text{dCS}}}{16 \pi} \int \mathrm{d}^4 x \sqrt{-g} \vartheta\mathcal{P}_{\text{dCS}},
\end{equation}
where $\mathcal{P}_{\text{dCS}} = (\star R)^{\mu\nu\alpha\beta}R_{\nu\mu\alpha\beta}$ is the Pontryagin curvature invariant and $(\star R)^{\mu\nu\alpha\beta} = \frac{1}{2}\epsilon^{\mu\nu\rho\sigma}R_{\rho\sigma}^{\ \ \ \alpha\beta}$ is the dual of the Riemann tensor.

Similarly to the parity-even scalar case, we can introduce $\zeta_{\text{dCS}} \sim O(\epsilon^2)$ \eqref{eq:bGR-parameter-scaling}. Then, stationary corrections to the background Kerr metric will appear at $O(\epsilon^4)$, while a nontrivial pseudoscalar background appears at $O(\epsilon^2)$. We can then expand the pseudoscalar field in powers of the mass-ratio as
\begin{align}
    \vartheta = \epsilon \theta , \quad \theta\sim O(1). 
\end{align} 

We now move to the analysis of $S_p[g,\vartheta]$ and its effective lagrangian \eqref{eq: effective lagrangian with theta - theta^2 - grad theta}.
By definition, $\vartheta$ transforms under a parity inversion as $\vartheta \rightarrow- \vartheta$. Hence, for $b_2 = k^{\mu}\nabla_{\mu}\vartheta$ to be a parity even scalar, $k^{\mu}$ needs to be parity odd. Using the CH theorem, together with Eq.~\eqref{eq: generic pseudovector out of u and Omega} and the identity $(\star \Omega)^{\alpha \sigma}\Omega_{\sigma \beta} \sim (\star \Omega)^{\mu\nu}\Omega_{\mu\nu} \delta^{\alpha}_{\beta}$, we have that the only leading-order parity-odd vector we can write down is
\begin{equation}
\label{eq:CH-pseudovector-leading}
    \hat k^{\mu} = \eta_1 (\star \hat\Omega)^{\mu\nu}u_{\nu}+\eta_2 (\star \hat\Omega)^{\alpha\beta}\hat\Omega_{\alpha\beta}u^{\mu} +O(\epsilon)\ ,
\end{equation}
where $\eta_{1,2}$ are adimensional constants.
Exploiting the electric-magnetic decomposition of the angular velocity introduced 
in Eq.~\eqref{eq: electric-magnetic decomposition of the angular velocity}, we write $(\star\hat\Omega)_{\mu\nu}\hat\Omega^{\mu\nu} \sim \hat E^{\mu} \hat B_{\mu}$. As $\hat E^{\mu} = O(\epsilon^2)$, the pseudoscalar satisfies $(\star\hat\Omega)_{\mu\nu}\hat\Omega^{\mu\nu}=O(\epsilon^2)$, and hence can be neglected.
Moreover, if one neglects spin-spin terms, it holds that (see, e.g., \cite{PhysRevD.73.104031}) $\hat\Omega_{\mu\nu} \sim \hat S_{\mu\nu}$. Hence, generalizing the definition of the spin (pseudo-)vector $(\star \hat S)^{\mu\nu}u_{\nu} = \hat S^{\mu}$ to $(\star \hat{ \mathcal{S}})^{\mu\nu}u_{\nu} = \hat{ \mathcal{S}}^{\mu}$, we have
\begin{equation}
\label{eq: dipole vector for pseudoscalar}
    \hat k^{\mu} = \xi \hat{\mathcal{S}}^{\mu}+O(\epsilon),
\end{equation}
for some dimensionless constant $\xi$. Notably, this is the same dipole vector considered in \cite{Loutrel:2018ydv}.

The scalar field appears explicitly in the particle Lagrangian also through $\vartheta$. The leading-order charge in the mass-ratio expansion is then whatever pseudoscalar $q(a_i)$ we can form using $g$ and the $a_i$ so that the Lagrangian is $\hat L(a_i,\vartheta,k^{\mu}\nabla_{\mu}\vartheta) = \hat L(a_i) + \epsilon \hat q(a_i)\theta + O(\epsilon^2)$. The only independent, nonzero choice is $\hat q(a_i) = \hat f(a_i)(\star \hat\Omega)_{\mu\nu}\hat\Omega^{\mu\nu}$, with $\hat f(a_i)$ a parity even scalar, as discussed in \ref{subsec:scalar-vs-pseudoscalar}. As $(\star \hat\Omega)_{\mu\nu}\hat \Omega^{\mu\nu} = O(\epsilon^2)$, the whole term proportional to $\theta$ appearing in the dimensionless $O(1)$ Lagrangian is $O(\epsilon^3)$, and hence negligible. 
At quadratic order in the scalar field, 
we can always write terms in the Lagrangian of the form $\sim \epsilon^2 \theta^2$ . However 
without a monopole moment, such a term 
is never relevant, as discussed 
in Sec.~\ref{sec: EMRI case - expanding field equations}.

This shows that the leading scalar multipole in the mass-ratio expansion of the secondary effective action is the scalar dipole 
vector $\hat{\mathcal{K}}^{\mu} = \hat{\mathcal{K}} \hat{\mathcal{S}}^{\mu}$, 
where $\hat{\mathcal{K}}$ is its 
constant amplitude.

Upon variation of the action \eqref{eq: general action - pseudoscalar} we 
obtain the field equation for $\vartheta$, 
neglecting $O(\epsilon^3)$ terms, 
\begin{equation}
    \label{eq: scalar field equation - mass-ratio expanded - dCS case}
   \hat\Box \theta =  - \epsilon \hat \alpha\hat{\mathcal{P}}^{(0)}_{\text{dCS}} + 16 \pi \epsilon \hat{\mathcal{K}}\hat \nabla_{\mu} \int \mathrm{d}\hat\tau \hat S^{\mu}  \frac{\hat\delta^{(4)}(x-z(\tau))}{\sqrt{-\hat g}}\ .
\end{equation} 
Equation~\eqref{eq: scalar field equation - mass-ratio expanded - dCS case} shows that  
the solution for $\theta$ is order 
$\sim O(\epsilon)$, at variance 
with our initial (conservative) 
assumption $\theta \sim O(1)$.
Such mass ratio suppression is consequence 
of the vanishing of the monopole term for the pseudoscalar. This implies that 
$\vartheta \sim \mathcal{O}(\epsilon^2)$ 
and hence that the contribution of 
the pseudoscalar to the scalar flux 
is order $\mathcal{O}(\epsilon^4)$, 
i.e. a $2\text{PA}$ effect that can 
be safely neglected.

Finally, following the procedure discussed in Sec.~\ref{sec:modifiedMPD-1PA}, while 
using the pseudovector $\hat k^{\mu}$ 
in Eq.~\eqref{eq: dipole vector for pseudoscalar}, we find the modified MPD equations for the pseudoscalar:
\begin{align}
\label{eq: modified linear momentum equation - dCS case}
    \frac{\mathrm{D}u^{\mu}}{\mathrm{d}\hat\tau} &= -\frac{1}{2}\epsilon \hat{R}^{\mu \alpha \beta \gamma}\hat{\mathcal{S}}_{\beta \gamma}u_{\alpha} 
    + \epsilon^2 \hat{\mathcal{K}} \nonumber \\ &
    \times \left( \hat{S}^{\rho}(g^{\mu\nu}+u^{\mu}u^{\nu})\hat{\nabla}_{\nu}\hat\nabla_{\rho}\theta -  \hat{\mathcal{S}}^{\mu}\frac{\mathrm{D}^2\theta}{\mathrm{d}\hat\tau^2} \right)\ ,
     \\ 
     \label{eq: modified spin tensor equation - dCS case}
    \frac{\mathrm{D}\hat{\mathcal{S}}^{\mu\nu}}{\mathrm{d}\hat\tau} & = 2 \epsilon \hat{\mathcal{K}} \hat{\mathcal{S}}^{[\mu}\hat\nabla_{\perp}^{\nu]}\theta \ .
\end{align}

The precession equation has a clear physical meaning: in the rest frame of the secondary, the spin precesses around the spatial direction of the axion vector $\grad_i \theta$. 

We note that, given 
$\theta \sim \mathcal{O}(\epsilon)$, 
the beyond GR terms in Eqns.~ \eqref{eq: modified linear momentum equation - dCS case}- \eqref{eq: modified spin tensor equation - dCS case} contribute at 
order $O(\epsilon^3)$, namely 
they affect the binary dynamics with 
$2\text{PA}$ (and higher) effects. 
The modified MPD equations for the 
pseudoscalar reduce 
to the GR case, and the correction to the effective particle Lagrangian \eqref{eq: effective lagrangian with theta - theta^2 - grad theta} vanishes at 1PA. 
This suggests that it is challenging to 
exploit GW observations of asymmetric binaries to constrain pseudoscalar-tensor theories, 
such as dCS gravity.
We note that there have  been previous studies on constraining dCS gravity with EMRIs~\cite{sopuerta2009extreme,pani2011gravitational,canizares2012testing,lui2026devoured}; however, they do not use the self-force expansion we use here~\cite{Maselli:2020zgv} and known constraints on dCS gravity~\cite{silva2021astrophysical} to assess at which post-adiabatic order the dCS contributions affect the EMRI.

Given deviations from GR for the pseudoscalar 
first appear at 2PA order, the 
self-force equations of motion for the secondary, at the 1PA order we consider here, are identical to GR.
Such equations can be found in Ref.~\cite{Mathews:2021rod}, or equivalently, 
they can be derived from the equations in \cref{sec:scalar-self-force-equations} by taking $\varphi^{(1)}=\varphi^{(2)}=0$, $\hat Q_0=0$, and $\hat Q_{\varphi\chi}=0$.

\section{Conclusions}

We have studied the dynamics of a spinning body in theories of gravity 
featuring a massless scalar field non-minimally coupled to gravity. We began by constructing the Lagrangian of a particle endowed with a scalar configuration and moving in an external background whose characteristic scale is much larger than the size of the body. To do so, we developed an approach for ``dressing up'' the particle action with an additional field, which can be seen as a generalization of Eardley's pioneering work, where the mass in the point-particle action was taken to be field-dependent~\cite{Eardley:1975fgi}. This lays the groundwork for a general treatment of the multipole moments 
of a body coupled to additional fields. We then used this approach to augment the GR action of a spinning body with the
scalar field. 

From the action we obtained, we derived the MPD equations of motion, which govern 
the acceleration and spin evolution of the body, as well as the field equations 
for the scalar field and the metric. We developed this formalism for both 
(parity-even) scalars and (parity-odd) pseudoscalars.
Working within an EFT framework, we then applied our formalism to extreme 
mass-ratio inspirals, exploiting the separation of scales naturally present in 
these systems to simplify the description of the orbital dynamics. In this setup, 
we expanded all relevant quantities in the mass ratio of the binary and 
computed corrections to GR up to first post-adiabatic (1PA) order. We have shown that higher mass or scalar multipoles do not contribute to 1PA order. Our results
extend and complete the framework developed in Refs.~\cite{Maselli:2020zgv,Spiers:2023cva}. 

For a scalar field, Eq.~\eqref{eq: effective lagrangian with phi - phi^2 - grad phi} presents the minimal form of the most general Lagrangian for a particle with spin and with a scalar monopole and dipole. The Lagrangian in Eq.~\eqref{1PAlangrangian} instead contains all terms and multipoles that contribute at 1PA. The corresponding equations of 
motion \eqref{eq: adimensional generalised MPD equation - spin tensor - final form - neglecting spin-spin}-\eqref{eq: equation for the four velocity - final form - neglecting spin-spin terms} 
and field equations \eqref{eq: scalar field equation - mass-ratio expanded}-\eqref{eq: gravitational field equations - mass-ratio expanded - neglecting spin-spin terms}  
contain new 1PA contributions, arising from couplings between 
the spin and the scalar field. These terms affect both the acceleration and the spin precession of 
the secondary. They are proportional to the scalar charge of the secondary and independent of its 
scalar dipole vector, which affects neither the dynamics of the body nor the fields at the order we consider.

For a pseudoscalar field, we were able to show using only the symmetries of the particle action that the secondary cannot have a pseudoscalar monopole, and that 
its dipole vector is proportional to the spin vector. This is consistent with the properties of stationary solutions in dCS gravity \cite{Yunes:2009hc}. Additionally, we showed that the dipole affects 
the binary dynamics and the GW emission only beyond 1PA order. Hence, up to this order 
the equations of motion and the field equations coincide with those of GR. This agrees with previous 
results in which the dipole vector was modeled as a pair of nearby scalar charges~\cite{Lestingi:2023ovn}.

The framework presented here opens several directions for future work. Since we restricted 
our analysis to massless fields, a natural extension is the study of spinning particles 
coupled to massive scalars, generalizing earlier treatments of non-rotating bodies~\cite{Barsanti:2022vvl}. 
In this case, the mass of the field introduces an additional length scale, as well as the 
possibility of forming bound states around the primary. This may require modifying the 
matching procedure on which the EMRI description is built, in order to account for a 
non-vacuum background.

Mapping the new quantities that enter the dynamics at 1PA order, 
such as the couplings between the spin and the scalar field, onto specific 
theories, such as scalar-Gauss-Bonnet gravity, would relate them directly to the 
fundamental couplings of those theories. This is a necessary step towards 
turning EMRI observations into constraints on specific extensions of GR.

Finally, within EMRI modeling, our results make it possible to quantify the 
relevance of the spin-dependent post-adiabatic effects identified here and to 
incorporate them into waveform models suitable for data analysis with future 
space-based detectors. Such models would allow us to assess both the 
improvement in constraints on beyond-GR parameters and the systematic 
biases introduced in tests of GR when these contributions are 
neglected~\cite{guo2024detecting,Barsanti:2026ulr}.

\begin{acknowledgments}

We thank S. Barsanti, H.L.H. Shum, and J. Purllant for helpful discussions, and R. Porto for clarifications on his paper during the early stages of this project.
TPS acknowledges partial support from the STFC Consolidated Grant nos. ST/V005596/1, ST/X000672/1, UKRI2492 and ST/C004235/1. 
AS~acknowledges funding from the European Union's Horizon Europe research and innovation programme under the Marie Sklodowska-Curie grant agreement no. 101199153. 

During the final stages of the project, the authors used Claude (Anthropic, Claude Opus 5.5) to check symbolic calculations and proofread the manuscript.
\end{acknowledgments}

\appendix

\section{Scalar-charge distribution from effective action}\label{app: Scalar-charge distribution from effective action}

Consider the general action 
\begin{equation}
    S = \int \mathrm{d}\tau L(\varphi,\nabla\varphi).
\end{equation}
Varying with respect to the scalar field, we have
\begin{align}
    \delta_{\varphi}S &= \int \mathrm{d}\tau \left(\frac{\partial L}{\partial \varphi}\delta \varphi + \frac{\partial L}{\partial \nabla_{\mu}\varphi}\nabla_{\mu}\delta \varphi \right) 
    \nonumber \\*
    &= \int \mathrm{d}\tau \int \mathrm{d}^4 x \sqrt{-g}\left(\frac{\partial L}{\partial \varphi}\delta \varphi + \frac{\partial L}{\partial \nabla_{\mu}\varphi}\nabla_{\mu}\delta \varphi \right)
    \nonumber \\*
    &\quad \times \frac{\delta^{(4)}(x-z(\tau))}{\sqrt{-g}}.
\end{align}
Integrating by parts the second term in the bracket gives
\begin{widetext}
\begin{equation}
   \delta_{\varphi}S = \int \mathrm{d}^4 x \sqrt{-g}  \left[ \int \mathrm{d}\tau\frac{\partial L}{\partial \varphi} \frac{\delta^{(4)}(x-z(\tau))}{\sqrt{-g}} - \nabla_\mu \int \mathrm{d}\tau \left( \frac{\partial L}{\partial \nabla_{\mu}\varphi} \frac{\delta^{(4)}(x-z(\tau))}{\sqrt{-g}}\right)  \right]\delta\varphi .
\end{equation}
One can then read off the source $\Sigma$:
\begin{equation}
    \Sigma =  -16 \pi \int \mathrm{d}\tau \frac{\partial L}{\partial \varphi}\frac{\delta^{(4)}(x-z(\tau))}{\sqrt{-g}} 
    + 16 \pi \nabla_{\mu}\int \mathrm{d}\tau \frac{\partial L}{\partial \nabla_\mu \varphi}\frac{\delta^{(4)}(x-z(\tau))}{\sqrt{-g}}\ .
\end{equation}
\end{widetext}
Introducing the effective charge $Q=\frac{\partial L}{\partial \varphi}$ and scalar dipole vector $K^{\mu}=\frac{\partial L}{\partial \nabla_\mu \varphi}$,
we now prove the invariance of the source under the transformation
\begin{align}
\label{eq:redundancy-rule}
    Q \rightarrow Q - \frac{\mathrm{d} f}{\mathrm{d}\tau}, \qquad K^{\mu} \rightarrow K^{\mu} - f u^{\mu}.
\end{align}
Applying the transformation, we see that in order for the invariance to hold, it must hold that
\begin{equation}
\label{eq:must-hold-invarince}
    \int \mathrm{d}\tau \frac{\mathrm{d}f}{\mathrm{d}\tau}\frac{\delta^{(4)}(x-z(\tau))}{\sqrt{-g}} = \nabla_{\mu} \int \mathrm{d}\tau f u^{\mu} \frac{\delta^{(4)}(x-z(\tau))}{\sqrt{-g}}.
\end{equation}
Consider the r.h.s.; by using $\nabla_{\mu}V^{\mu} = (\sqrt{-g})^{-1}\partial_{\mu}(\sqrt{-g}V^{\mu})$ and that $\partial_\mu (f(\tau)u^{\mu})=0$, we get 
\begin{equation}
    \begin{split}
    \nabla_{\mu} \! \int \! \mathrm{d}\tau f u^{\mu} \frac{\delta^{(4)}(x-z(\tau))}{\sqrt{-g}} &= \frac{1}{\sqrt{-g}}\!\int \!\mathrm{d}\tau f u^{\mu } \\
    &\quad\times\partial_\mu \delta^{(4)}(x-z(\tau)).
    \end{split}
\end{equation}
Now note that $\frac{\partial}{\partial x^{\mu}}\delta^{(4)}(x-z(\tau)) = - \frac{\partial}{\partial z^{\mu}}\delta^{(4)}(x-z(\tau))$. Hence, $ u^{\mu } \partial_\mu \delta^{(4)}(x-z(\tau)) = - \frac{\mathrm{d}}{\mathrm{d}\tau}\delta^{(4)}(x-z(\tau)) $. Finally, integrating by parts gives us the l.h.s of \eqref{eq:must-hold-invarince}, showing the invariance of the source under \eqref{eq:redundancy-rule}.

\section{Dipole correction to spin-linear momentum relations}\label{app: Dipole correction to spin-linear momentum relations}

Let us start from the Lagrangian $L(g,u,\Omega,\varphi,\nabla_{\mu}\varphi)$. As $L$ is a scalar, an infinitesimal coordinate transformation $x^{\mu} \rightarrow x^{\mu} + \delta x^\mu, \ \delta x^{\mu} = \epsilon^{\mu}_{\nu}x^{\nu}$ yields $\delta L=0$. On the other hand, the variation of $L$ can be written as
\begin{align}
    \delta L &= \frac{\partial L}{\partial g^{\mu\nu}}\Bigg|_\Omega\delta g^{\mu\nu} + \frac{\partial L}{\partial u^{\mu}} \delta u^{\mu} + \frac{\partial L}{\partial \Omega_{\mu\nu}}\delta \Omega_{\mu\nu} + \frac{\partial L}{\partial \nabla_\mu \varphi}\delta \nabla_\mu \varphi
    \nonumber \\* 
    & = \mathcal{P}_{\mu} \delta u^{\mu} + \frac{1}{2}\mathcal{S}^{\mu\nu}\delta \Omega_{\mu\nu} + \frac{\partial L}{\partial g^{\mu\nu}}\Bigg|_\Omega\delta g^{\mu\nu} + \frac{\partial L}{\partial \nabla_\mu \varphi}\delta \nabla_\mu \varphi
    ,
\end{align}
where no $\delta \varphi$ appears as $\varphi$ is a scalar, and with $|_{\Omega}$ implying that $\Omega$ is being held fixed. Under infinitesimal coordinate transformations vectors and tensors transform as $\delta V^{\mu} = \epsilon^{\mu}_\nu V^{\nu} $ and $\delta X_{\mu\nu} = -\epsilon^\rho_\mu X_{\rho\nu}-\epsilon^{\sigma}_{\nu}X_{\mu\sigma}$, respectively.
Using the above transformations, we have 
\begin{equation}
    \delta L = \epsilon^{\nu \mu}\left( \mathcal{P}_{\nu} u_{\mu} - \mathcal{S}_{\mu}^{\ \  \rho}\Omega_{\nu\rho}+ 2 \frac{\partial L}{\partial g^{\mu\nu}}\Bigg|_\Omega - \frac{\partial L}{\partial \nabla^{\mu}\varphi}\nabla_{\nu}\varphi \right).
\end{equation}
Introducing the scalar dipole vector $K^{\mu} = \frac{\partial L}{\partial \nabla_{\mu}\varphi}$ and using $\delta L = 0 $, we find the relation
\begin{equation}
\label{eq:magic-relation}
    2 \frac{\partial L}{\partial g^{\mu\nu}}\Bigg|_\Omega + \mathcal{S}_{\mu\rho}\Omega^{\rho}_{\ \nu} = -\mathcal{P}_{\nu}u_{\mu} + K_{\mu}\nabla_{\nu}\varphi.
\end{equation}

\section{Deriving pole-dipole stress-energy tensor from effective action}\label{app: Deriving pole-dipole stress-energy tensor from effective action}

We vary $S_p$ with respect to $g^{\mu\nu}$ at fixed worldline $z^\mu(\tau)$, parameter $\tau$, and at fixed $\varphi$ and hence fixed $\nabla_\mu\varphi$. Thus, $\delta_g u^\mu=\delta_g\varphi=\delta_g\nabla_\mu\varphi=0$. The variation of the tetrad reads:
\begin{equation}
\label{eq:D1}
\delta_g e^I_\mu=\frac{1}{2} \delta g_{\mu\nu} e^{I\nu}, \qquad \delta g_{\mu\nu} = -g_{\mu\alpha} g_{\nu\beta} \delta g^{\alpha\beta},
\end{equation}
which satisfies $\delta_g(\eta_{IJ} e^I_\mu e^J_\nu) = \delta g_{\mu\nu}$. 

We now want to compute the variation of the angular velocity $\Omega_{\mu\nu}$.
Write Eq.~\eqref{eq: angular velocity and tetrad} as $\Omega_{\mu\nu}=e^I_\mu u^\lambda \left(\partial_\lambda e_{\nu I} - \Gamma^\rho_{\lambda\nu}e_{\rho I}\right)$. Use Eq.~\eqref{eq:D1}, which gives $\delta e_{\nu I}=\frac{1}{2} \delta g_\nu^{\sigma}e_{\sigma I}$ with $\delta g_\nu^{\sigma}= g^{\sigma\rho}\delta g_{\nu\rho}$, together with the standard variation of the connection
\begin{equation}
\label{eq:D2}
\delta\Gamma_{\mu\lambda\nu}= g_{\mu\rho} \delta\Gamma^\rho{}_{\lambda\nu} = \frac{1}{2}\left(\nabla_\lambda\delta g_{\mu\nu} + \nabla_\nu\delta g_{\mu\lambda} - \nabla_\mu\delta g_{\lambda\nu}\right).
\end{equation}
One finds
\begin{align}
\delta_g\Omega_{\mu\nu}
=&\delta e^I{}_\mu \frac{\mathrm{D}e_{\nu I}}{\mathrm{d}\tau}+e^I{}_\mu \frac{\mathrm{D} \delta e_{\nu I}}{\mathrm{d}\tau}-u^\lambda \delta\Gamma_{\mu\lambda\nu} 
\nonumber\\
=&\frac{1}{2} \delta g_\mu{}^{\alpha}\,\Omega_{\alpha\nu}+\left(\frac{1}{2} u^\lambda\nabla_\lambda\delta g_{\mu\nu}-\frac{1}{2}\delta g_\nu{}^{\alpha} \Omega_{\alpha\mu}\right)
\nonumber \\ 
&-\frac{1}{2} u^\lambda\left(\nabla_\lambda\delta g_{\mu\nu}+\nabla_\nu\delta g_{\mu\lambda}-\nabla_\mu\delta g_{\lambda\nu}\right)\nonumber\\
=& u^\lambda\nabla_{[\mu}\delta g_{\nu]\lambda}+g_{\alpha[\mu}\Omega_{\nu]\beta} \delta g^{\alpha\beta},
\label{eq:D3}
\end{align}
which is Eq.\eqref{eq:variation-angular-velocity}.

Since $\tau$ and $u^\mu$ are metric independent,
\begin{equation}
\delta_g S_p=\int\mathrm{d}\tau\left[\frac{\partial L}{\partial g^{\mu\nu}}\Bigg|_\Omega\delta g^{\mu\nu}+\frac{1}{2} \mathcal{S}^{\beta\alpha} \delta_g\Omega_{\beta\alpha}\right],
\label{eq:D4}
\end{equation}
with $\partial L/\partial g^{\mu\nu}|_\Omega$ given by Eq.~\eqref{eq:lagrangian-metric-derivative}. The algebraic term in Eq.~\eqref{eq:D3} gives $\frac{1}{2}\mathcal{S}_{(\mu|\alpha}\Omega^\alpha{}_{\nu)} \delta g^{\mu\nu}$. For the derivative term we use $\nabla_\beta\delta g_{\alpha\lambda}=-g_{\alpha\rho}g_{\lambda\sigma}\nabla_\beta\delta g^{\rho\sigma}$:
\begin{equation}
\frac{1}{2}\mathcal{S}^{\beta\alpha}u^\lambda \nabla_\beta\delta g_{\alpha\lambda} = -\frac{1}{2} \mathcal{S}^{\beta}{}_{\rho} u_\sigma \nabla_\beta\delta g^{\rho\sigma}.
\label{eq:D5}
\end{equation}
Writing the worldline integrals as $\int\mathrm{d}^4 x \sqrt{-g}\int\mathrm{d}\tau (\ldots) \delta^{(4)}(x-z(\tau))/\sqrt{-g}$ and integrating Eq.~\eqref{eq:D5} by parts in $x$, we get
\begin{align}
\delta_g S_p &= \int\mathrm{d}^4x\sqrt{-g} \delta g^{\mu\nu} \left(\int\mathrm{d}\tau \frac{\delta^{(4)}(x-z(\tau))}{\sqrt{-g}} \left[\frac{\partial L}{\partial g^{\mu\nu}} \Bigg|_\Omega \right. \right.
\nonumber \\ & \left. \left.
+\frac{1}{2} \mathcal{S}_{(\mu|\alpha}\Omega^\alpha{}_{\nu)}\right]+\frac{1}{2} \nabla^\rho\!\int\mathrm{d}\tau \frac{\delta^{(4)}(x-z(\tau))}{\sqrt{-g}} \mathcal{S}_{\rho(\mu}u_{\nu)}\right).
\label{eq:D6}
\end{align}
Using Eq.~\eqref{eq:stressenergycp}, $T^p_{\mu\nu}=-(16\pi/\sqrt{-g}) \delta S_p/ \delta g^{\mu\nu}$, this gives
\begin{align}
T^p_{\mu\nu}=& -16\pi\int\mathrm{d}\tau \frac{\delta^{(4)}(x-z(\tau))}{\sqrt{-g}} \left[\frac{\partial L}{\partial g^{\mu\nu}}\Bigg|_\Omega+\frac{1}{2} \mathcal{S}_{(\mu|\alpha}\Omega^\alpha{}_{\nu)}\right]
\nonumber \\
&
-8\pi \nabla^\rho\!\int\mathrm{d}\tau \frac{\delta^{(4)}(x-z(\tau))}{\sqrt{-g}} \mathcal{S}_{\rho(\mu}u_{\nu)} ,
\label{eq:D7}
\end{align}
i.e., Eq.~\eqref{eq:stressenergy-beforemagicformula}. We can now simplify the bracket using the identity \eqref{eq:magic-relation} symmetrized
\begin{equation}
\frac{\partial L}{\partial g^{\mu\nu}}\Bigg|_\Omega + \frac{1}{2}\mathcal{S}_{(\mu|\alpha}\Omega^\alpha{}_{\nu)} = \frac{1}{2}\left[-\mathcal{P}_{(\mu}u_{\nu)}+K_{(\mu}\nabla_{\nu)}\varphi\right],
\label{eq:D8}
\end{equation}
so that:
\begin{align}
\label{eq:D9}
T^p_{\mu\nu}=& 8\pi\int\mathrm{d}\tau \frac{\delta^{(4)}(x-z(\tau))}{\sqrt{-g}} \left[\mathcal{P}_{(\mu}u_{\nu)}-K_{(\mu}\nabla_{\nu)}\varphi\right]
\nonumber \\ &
-8\pi \nabla^\rho\!\int\mathrm{d}\tau \frac{\delta^{(4)}(x-z(\tau))}{\sqrt{-g}} \mathcal{S}_{\rho(\mu}u_{\nu)} ,
\end{align}
which is Eq.~\eqref{eq: stress-energy tensor with scalar dipole correction}.

\section{Modified MPD equations from effective action}\label{app: Modified MPD equations from effective action}

Here we derive the modified MPD equations \eqref{eq: generalised MPD equation - linear momentum}, \eqref{eq: generalised MPD equation - spin tensor} by varying $S_p = \int\mathrm{d}\tau L(a_i,\varphi,k^\mu \nabla_\mu \varphi)$ with respect to the worldline $z^\mu(\tau)$ and the tetrad, at fixed metric and scalar field. We work at an arbitrary point $z(\tau_0)$ on the worldline in Riemann normal coordinates centered there, and then show that the equations we end up with are tensorial. In these coordinates $\Gamma^\mu_{\alpha\beta}(z(\tau_0))=0$ and $\partial_\alpha g_{\mu\nu}(z(\tau_0))=0$.

The variation of $L$ with respect to the worldline, Eq.~\eqref{eq: final delta-x L}, splits into a part with the same form as in GR and the terms involving the scalar field:
\begin{align}
\delta_z L = & \left[\frac{\partial L}{\partial g_{\mu\nu}} \delta_z g_{\mu\nu} + \mathcal{P}_\mu \delta_z u^\mu+\frac{1}{2} \mathcal{S}^{\mu\nu} \delta_z\Omega_{\mu\nu}\right]
\nonumber \\ &
+Q \delta_z\varphi+K^\mu \delta_z\nabla_\mu\varphi ,
\label{eq:FinalApp1}
\end{align}
with $\mathcal{P}^\mu$ and $\mathcal{S}^{\mu\nu}$ defined in Eqs.~\eqref{eq: generalized linear momentum - main tex}, \eqref{eq: generalized spin tensor - main tex}. Any dependence of $L$ on $(g,u,\Omega)$ through $b_2 = k^{\mu}\nabla_{\mu}\varphi$ is already contained in these definitions. The bracket is therefore formally identical to the GR case of Ref.~\cite{PhysRevD.73.104031}, with $p^\mu\to\mathcal{P}^\mu$ and $S^{\mu\nu}\to\mathcal{S}^{\mu\nu}$. After integration by parts, it contributes to the integrand of $\delta S_p$ at $z(\tau_0)$ the GR expression
\begin{equation}
\delta z^\alpha\left[-\frac{\mathrm{D}\mathcal{P}_\alpha}{\mathrm{d}\tau}-\frac{1}{2} R_{\alpha\beta\mu\nu} u^\beta\mathcal{S}^{\mu\nu}\right].
\label{eq:FinalApp2}
\end{equation}
For the scalar-field terms, $\delta_z\varphi = \delta z^\alpha\partial_\alpha\varphi$ and $\delta_z\nabla_\mu\varphi = \delta z^\alpha\partial_\alpha\partial_\mu\varphi$. Since $\partial_\alpha\partial_\mu\varphi = \nabla_\alpha\nabla_\mu\varphi + \Gamma^\lambda{}_{\alpha\mu}\partial_\lambda\varphi$ and the Christoffel symbols vanish at $z(\tau_0)$, these terms give
\begin{equation}
Q \delta_z\varphi + K^\mu \delta_z\nabla_\mu\varphi = \delta z^\alpha\left[Q \nabla_\alpha\varphi + K^\mu\nabla_\alpha\nabla_\mu\varphi\right].
\label{eq:FinalApp3}
\end{equation}
Combining Eqs.~\eqref{eq:FinalApp2} and \eqref{eq:FinalApp3} and requiring $\delta S_p=0$ for arbitrary $\delta z^\alpha$ gives
\begin{equation}
\frac{\mathrm{D}\mathcal{P}^\mu}{\mathrm{d}\tau} = -\frac{1}{2} R^{\mu\alpha\beta\gamma} \mathcal{S}_{\beta\gamma} u_\alpha + Q \nabla^\mu\varphi + K^\rho\nabla_\rho\nabla^\mu\varphi .
\label{eq:FinalApp4}
\end{equation}
Every term in Eq.~\eqref{eq:FinalApp4} is a tensor evaluated at $z(\tau_0)$, so the equation holds in any coordinate system. Since $z(\tau_0)$ is arbitrary, it holds along the whole worldline. This is the modified MPD equation for the linear momentum, Eq.~\eqref{eq: generalised MPD equation - linear momentum}.

In general coordinates, neither the bracket in Eq.~\eqref{eq:FinalApp1} nor $K^\mu\partial_\alpha\partial_\mu\varphi$ is separately covariant. In GR, the Christoffel terms of the bracket cancel among themselves. When $L$ also depends on $\nabla_\mu\varphi$, the identity \eqref{eq:magic-relation} shows that they leave a residual $-K^\mu\Gamma^\lambda_{\alpha\mu}\partial_\lambda\varphi \delta z^\alpha$, which combines with $K^\mu\partial_\alpha\partial_\mu\varphi \delta z^\alpha$ into $K^\mu\nabla_\alpha\nabla_\mu\varphi \delta z^\alpha$. Normal coordinates avoid carrying all these Christoffel symbols.

The variation with respect to the tetrad is identical to the GR case (for which we refer to \cite{PhysRevD.73.104031}), with $S^{\mu\nu} \rightarrow \mathcal{S}^{\mu\nu}$, as both terms depending explicitly on the scalar field are zero when varied with respect to the tetrad. This gives the first equality of the modified MPD equation for the spin tensor, Eq.~\eqref{eq: generalised MPD equation - spin tensor}.

\bibliography{bibliography}
\end{document}